\documentclass[a4paper,UKenglish,cleveref,autoref,thm-restate]{lipics-v2021}

\usepackage{todonotes}

\usepackage[createShortEnv,conf={end,link to proof,restate}]{proof-at-the-end}

\newcommand{\skipnoindent}{\noindent}

\newcommand{\arity}{\#}
\newcommand{\arityof}[1]{\arity{#1}}

\newcommand{\cardof}[1]{\mathrm{card}({#1})}

\newcommand{\univ}{\mathsf{U}}

\newcommand{\isdef}{\stackrel{\scalebox{.4}{$\mathsf{def}$}}{=}}

\newcommand{\interv}[2]{[{#1}..{#2}]}

\newcommand{\set}[1]{\{ {#1} \}}
\newcommand{\mset}[1]{\{\!\!\{{#1}\}\!\!\}}

\newcommand{\pow}[1]{\mathrm{pow}({#1})}
\newcommand{\mpow}[1]{\mathrm{mpow}({#1})}

\newcommand{\finsubseteq}{\subseteq_{\mathit{fin}}}

\newcommand{\fsignature}{\mathbb{F}}
\newcommand{\sign}{\mathcal{F}}
\newcommand{\termsof}[1]{\mathcal{T}({#1})}
\newcommand{\relations}{\mathbb{R}}
\newcommand{\constants}{\mathbb{C}}
\newcommand{\rels}{\mathcal{R}}
\newcommand{\csts}{\mathcal{C}}

\newcommand{\algof}[1]{\mathcal{#1}}
\newcommand{\universeOf}[1]{\mathsf{#1}}

\newcommand{\arel}{\mathsf{r}}

\newcommand{\aarel}{\mathsf{a}}
\newcommand{\abrel}{\mathsf{b}}
\newcommand{\acrel}{\mathsf{c}}

\newcommand{\acst}{\mathsf{c}}

\newcommand{\arrow}[2]{\xrightarrow{{\scriptscriptstyle #1}}_{{\scriptstyle #2}}}

\newcommand{\auto}[2]{\mathcal{A}_{
    {#1}
    \ifthenelse{\equal{#2}{}}{}{,{#2}}
}}

\newcommand{\autsat}[2]{\mathcal{A}^{I}_{
    {#1}
    \ifthenelse{\equal{#2}{}}{}{,{#2}}
}}

\newcommand{\autcut}[2]{\mathcal{A}^{\scriptscriptstyle\mathsf{cut}}_{
    {#1}
    \ifthenelse{\equal{#2}{}}{}{,{#2}}
}}

\newcommand{\autcst}[2]{\mathcal{A}^{\scriptscriptstyle\mathsf{cst}}_{
    {#1}
    \ifthenelse{\equal{#2}{}}{}{,{#2}}
  }
}

\makeatletter
\newcommand*{\da@rightarrow}{\mathchar"0\hexnumber@\symAMSa 4B }
\newcommand*{\da@leftarrow}{\mathchar"0\hexnumber@\symAMSa 4C }
\newcommand*{\xdashrightarrow}[2][]{%
  \mathrel{%
    \mathpalette{\da@xarrow{#1}{#2}{}\da@rightarrow{\,}{}}{}%
  }%
}
\newcommand{\xdashleftarrow}[2][]{%
  \mathrel{%
    \mathpalette{\da@xarrow{#1}{#2}\da@leftarrow{}{}{\,}}{}%
  }%
}
\newcommand*{\da@xarrow}[7]{%
  \sbox0{$\ifx#7\scriptstyle\scriptscriptstyle\else\scriptstyle\fi#5#1#6\m@th$}%
  \sbox2{$\ifx#7\scriptstyle\scriptscriptstyle\else\scriptstyle\fi#5#2#6\m@th$}%
  \sbox4{$#7\dabar@\m@th$}%
  \dimen@=\wd0 %
  \ifdim\wd2 >\dimen@
    \dimen@=\wd2 %
  \fi
  \count@=2 %
  \def\da@bars{\dabar@\dabar@}%
  \@whiledim\count@\wd4<\dimen@\do{%
    \advance\count@\@ne
    \expandafter\def\expandafter\da@bars\expandafter{%
      \da@bars
      \dabar@
    }%
  }%
  \mathrel{#3}%
  \mathrel{%
    \mathop{\da@bars}\limits
    \ifx\\#1\\%
    \else
      _{\copy0}%
    \fi
    \ifx\\#2\\%
    \else
      ^{\copy2}%
    \fi
  }%
  \mathrel{#4}%
}
\makeatother

\newcommand{\struc}{\sigma}

\newcommand{\allstof}[4]{
  \Omega_{#1}^{{#2}/{#3}}
  \ifthenelse{\equal{#4}{}}{}{({#4})}
}

\newcommand{\strucof}[2]{\mathcal{S}({#2})}
\newcommand{\strucs}{\mathbf{S}}

\newcommand{\astruc}{\mathsf{S}}

\newcommand{\restrict}[2]{
  \mathsf{restr}_{\scriptscriptstyle{{#1} \ifthenelse{\equal{#2}{}}{}{,}{#2}}}
}

\newcommand{\rename}[2]{
  {\mathsf{rename}^{\scriptscriptstyle{{#1}}}_{\scriptscriptstyle{{#2}}}}
}

\newcommand{\urename}[2]{
  \overline{\mathsf{rename}}_{\scriptscriptstyle{{#1} \ifthenelse{\equal{#2}{}}{}{,}{#2}}}
}

\newcommand{\forget}[2]{
  {\mathsf{forget}^{\scriptscriptstyle{{#1}}}_{\scriptscriptstyle{{#2}}}}
}

\newcommand{\uforget}[2]{
  \overline{\mathsf{forget}}_{\scriptscriptstyle{{#1} \ifthenelse{\equal{#2}{}}{}{,}{#2}}}
}

\newcommand{\relab}[2]{
  \mathsf{relab}_{\scriptscriptstyle{{#1} \ifthenelse{\equal{#2}{}}{}{,}{#2}}}
}

\newcommand{\addedge}[2]{
  \mathsf{add}_{\scriptscriptstyle{{#1} \ifthenelse{\equal{#2}{}}{}{,}{#2}}}
}

\newcommand{\fuse}[3]{\mathsf{F}({#2},{#3})}
\newcommand{\fclose}[2]{\mathsf{F}^*_{#1}({#2})}

\newcommand{\comp}{*}

\newcommand{\pop}[2]{\oplus_{\scriptscriptstyle{{#1}
      \ifthenelse{\equal{#2}{}}{}{,{#2}}}}
}

\newcommand{\upop}[2]{\overline{\oplus}_{\scriptscriptstyle{{#1}
      \ifthenelse{\equal{#2}{}}{}{,{#2}}}}
}

\newcommand{\vrpop}[2]{\uplus_{\scriptscriptstyle{{#1}
      \ifthenelse{\equal{#2}{}}{}{,{#2}}}}
}

\newcommand{\mso}{\textsf{MSO}}
\newcommand{\msof}[2]{$\mathsf{MSO}\ifthenelse{\equal{#1}{}\AND\equal{#2}{}}{}{({#1},{#2})}$}

\newcommand{\slr}{\textsf{SLR}}

\newcommand{\sintfusion}[2]{\widetilde{\mathtt{IF}}({#1}\ifthenelse{\equal{#2}{}}{}{,{#2}})}

\newcommand{\lang}[1]{\mathcal{L}({#1})}

\newcounter{index}

\newcommand{\gaifmanof}[1]{\mathsf{Gaif}({#1})}

\newcommand{\graph}{G}

\newcommand{\graphsof}[2]{\mathcal{G}\ifthenelse{\equal{#1}{}\AND\equal{#2}{}}{}{({#1},{#2})}}
\newcommand{\vertices}{V}
\newcommand{\vertof}[1]{\vertices_{\scriptscriptstyle{#1}}}
\newcommand{\edges}{E}
\newcommand{\edgeof}[1]{\edges_{\scriptscriptstyle{#1}}}

\newcommand{\slabs}{\mathcal{C}}

\newcommand{\sorts}{\Sigma}
\newcommand{\argof}[1]{\alpha({#1})}
\newcommand{\sortof}[1]{\rho({#1})}

\newcommand{\grammar}{\Gamma}

\newcommand{\hr}{$\mathsf{HR}$}
\newcommand{\vr}{$\mathsf{VR}$}

\newcommand{\colors}{\mathfrak{C}}
\newcommand{\acolor}{\gamma}

\newcommand{\twof}[1]{\mathrm{tw}({#1})}

\newcommand{\extfusionone}[2]{\mathsf{F}_1({#1},{#2})}

\newcommand{\absextfusionone}[2]{\mathsf{f}_1^{\scriptscriptstyle{\sharp}}({#1},{#2})}
\newcommand{\absreachfusionone}[1]{\mathsf{f}_1^{\scriptscriptstyle\sharp*}({#1})}

\newcommand{\kabsextfusionone}[3]{\mathsf{f}_1^{\scriptscriptstyle\sharp{#1}}({#2},{#3})}
\newcommand{\kabsreachfusionone}[2]{\mathsf{f}_1^{\scriptscriptstyle\sharp{#1}*}\ifthenelse{\equal{#2}{}}{}{({#2})}}

\newcommand{\mcolabs}[1]{{#1}^{\sharp}}
\newcommand{\kmcolabs}[2]{{#2}^{\scriptscriptstyle\sharp{#1}}}

\newcommand{\allcols}{\Gamma}
\newcommand{\bluecols}{\allcols^{blue}}
\newcommand{\greencols}{\allcols^{green}}
\newcommand{\redcols}{\allcols^{red}}

\newcommand{\colorof}[2]{\mathsf{col}_{\ifthenelse{\equal{#1}{}}{{#2}}{{#1}}}}

\newcommand{\aleaf}{\lambda}
\newcommand{\leafof}[1]{\aleaf({#1})}

\newcommand{\hastype}{\tau}
\newcommand{\typeof}[1]{\hastype({#1})}
\newcommand{\amode}{\mu}
\newcommand{\bluetype}{\mathsf{B}}
\newcommand{\greentype}{\mathsf{G}}
\newcommand{\redtype}{\mathsf{R}}

\newcommand{\nodelab}[1]{\mathsf{lab}(#1)}
\newcommand{\labelop}[4]{\mathsf{label}({#1},{#2},{#3},{#4})}
\newcommand{\joinop}[4]{\mathsf{join}({#1},{#2},{#3},{#4})}
\newcommand{\appendop}[5]{\mathsf{append}({#1},{#2},{#3},{#4},{#5})}

\title{Iterating Non-Aggregative Structure Compositions}

\author{Marius Bozga}{Univ. Grenoble Alpes, CNRS, Grenoble INP, VERIMAG, 38000, France}{Marius.Bozga@univ-grenoble-alpes.fr}{https://orcid.org/0000-0003-4412-5684}{}
\author{Radu Iosif}{Univ. Grenoble Alpes, CNRS, Grenoble INP, VERIMAG, 38000, France}{Radu.Iosif@univ-grenoble-alpes.fr}{https://orcid.org/0000-0003-3204-3294}{}
\author{Florian Zuleger}{Institute of Logic and Computation, Technische Universit\"{a}t Wien, Austria}{Florian.Zuleger@tuwien.ac.at}{https://orcid.org/0000-0003-1468-8398}{}
\authorrunning{M. Bozga and R. Iosif and F. Zuleger} 

\ccsdesc[100]{\textcolor{red}{•Theory of computation~Logic~Graph grammars}} 

\keywords{Hyperedge replacement,~Tree-width} 

\nolinenumbers 

\begin{document}

\maketitle

\begin{abstract}
  An aggregative composition is a binary operation obeying the
  principle that the whole is determined by the sum of its parts. The
  development of graph algebras, on which the theory of formal graph
  languages is built, relies on aggregative compositions that behave
  like disjoint union, except for a set of well-marked interface
  vertices from both sides, that are joined. The same style of
  composition has been considered in the context of relational
  structures, that generalize graphs and use constant symbols to label
  the interface.

  In this paper, we study a non-aggregative composition operation,
  called \emph{fusion}, that joins non-deterministically chosen
  elements from disjoint structures. The sets of structures obtained
  by iteratively applying fusion do not always have bounded
  tree-width, even when starting from a tree-width bounded set.
  First, we prove that the problem of the existence of a bound on the
  tree-width of the closure of a given set under fusion is decidable,
  when the input set is described inductively by a finite
  \emph{hyperedge-replacement} (HR) grammar, written using the
  operations of aggregative composition, forgetting and renaming of
  constants. Such sets are usually called \emph{context-free}.
  Second, assuming that the closure under fusion of a context-free set
  has bounded tree-width, we show that it is the language of an
  effectively constructible HR grammar. A possible application of the
  latter result is the possiblity of checking whether all structures
  from a non-aggregatively closed set having bounded tree-width
  satisfy a given monadic second order logic formula.
\end{abstract}

\section{Introduction}

The \emph{tree-width} of a graph is a numerical measure of how
“tree-like” the graph is.  This notion extends naturally to the
relational structures, used to define the semantics of classical first
and second-order logic. Relational structures generalize a broad range
of graph-like objects, such as edge-labeled graphs, hypergraphs, and
multi-edge graphs. Tree-width plays a foundational role in logic and
verification. Courcelle’s theorem~\cite{CourcelleI} shows that
\emph{Monadic Second-order Logic} (\mso) is decidable on classes of
structures having bounded tree-width, while Seese’s
theorem~\cite{Seese91} asserts that unbounded tree-width leads to
undecidability of \mso{} theories. Thus, proving that a given class of
structures has bounded tree-width is tantamount for establishing the
decidability of logical theories for that class.

In principle, one is interested in reasoning about infinite families
of structures. These families are typically generated inductively from
a finite set of basic building blocks, using operations such as
composition and renaming. These operations are usually formalized by
the Hyperedge Replacement (\hr) algebra introduced by
Courcelle~\cite{courcelle_engelfriet_2012}. The principle of inductive
definition is captured by the notion of a \emph{context-free} set,
i.e., the language of a finite grammar written using \hr{} operations,
or equivalently, the set of evaluations of a set of ground \hr{} terms
recognized by a tree automaton.

For both graphs and relational structures, \hr{} algebras are built on
\emph{aggregative composition} operations, in which the whole is
determined by the sum of its parts. These compositions are typically
defined as the disjoint union of the arguments, where the elements
designated by shared constants on both sides are joined together. In
other words, aggregative composition preserves the identity of the
substructures while merging them at known interface
points. Importantly, context-free sets of graphs and relational
structures defined by grammars based on such
bounded-interface\footnote{Vertex-replacement
algebras~\cite{courcelle_engelfriet_2012} using disjoint union and
edge addition between arbitrarily large interfaces may produce sets of
unbounded tree-width.} aggregative compositions have bounded
tree-width.

In this paper, we investigate a more flexible but less controlled
operation, called \emph{non-aggregative fusion}. Fusion allows
elements from two disjoint structures to be merged
nondeterministically, even when they are not marked by constants.  To
maintain a certain level of semantic coherence, we require fusion to
be constrained by a \emph{coloring} discipline: elements can only be
joined if their colors (i.e., sets of designated unary relations) are
disjoint. This idea stems from existing work in the area of reasoning
about the correctness of systems with dynamically reconfigurable
connectivity, such as distributed
protocols~\cite{AhrensBozgaIosifKatoen21} or pointer structures with
aliasing~\cite{OHearnReynoldsYang01}.

Due to its nondeterministic nature, fusion can cause a dramatic shift
in structure: even if a set $\strucs$ consists of structures having
bounded tree-width, by taking the closure of $\strucs$ under fusion
one may introduce infinitely many structures of unbounded tree-width.
This phenomenon raises two natural and fundamental
questions: \begin{enumerate}
\item Given a context-free set of relational structures, does the
  closure of this set under fusion have bounded tree-width ?
\item If the answer to the above question is yes, is this closure
  again a context-free set ? 
  %
\end{enumerate}
The result this paper is that both questions have a positive answer
(Theorem~\ref{thm:main}): \begin{enumerate}
\item The existence of a bound on the tree-width of the closure by
  fusion of a context-free set is a decidable problem.
\item If the fusion-closure of context-free set has bounded
  tree-width, then it is the language of an effectively constructible
  context-free grammar, that uses only aggregative composition.
\end{enumerate}
These results provide tools for reasoning about nondeterministic
structural iteration. For instance, one can check \mso{} properties
over the tree-width bounded fusion-closure of a context-free set,
thereby extending algorithmic verification techniques to a broader
class of systems. We sketch below two possible application domains
that have motivated our work.

\noindent\emph{Separation Logic of Relations (\slr)} A key motivation
for studying non-aggregative fusion comes from \slr, a generalization
of classical Separation Logic to relational structures. This logic has
been first considered for relational databases and object-oriented
languages~\cite{10.1007/978-3-540-27864-1_26}.  More recently, \slr{}
(combined with inductive definitions~\cite{Concur23}) has been
proposed as an assertion language for the verification of distributed
reconfigurable systems~\cite{AhrensBozgaIosifKatoen21}.  Here, the
\emph{separating conjunction} $\phi * \psi$ means that the models of
two formul{\ae} $\phi$ and $\psi$ must not have overlapping
interpretations of the same relation symbol.

A subtle but crucial point is that, while the separating conjunction
enforces \emph{disjointness} of the tuples that interpret a relation
symbol, i.e., that tuples cannot overlap in \emph{all} positions, the
tuples may overlap in \emph{some} positions. However, by the
disjointness of $*$, such overlapping is only possible if the
variables at these positions do not occur within the same unary
relation symbol \footnote{For each unary relation symbol $\arel$ in
the alphabet, the \slr{} formula $\arel(x) * \arel(y)$ entails $x\neq
y$.}.  From a semantic point of view, this behavior corresponds
precisely to our notion of fusion: joining elements of two separate
structures is allowed, as long as their sets of unary relation labels
are disjoint.  Thus, fusion abstracts the semantics of \slr{}, where
aliasing is controlled implicitly by colors (i.e., sets of relation
symbols) and the semantics of the separating conjunction.  Moreover,
considering inductive definitions on top of the basic \slr{} logic is
akin to considering context-free sets generated by recursive grammars
here.

\noindent\emph{Chemical and Biological Systems} We believe that
fusion-like operations naturally arise in the modeling of chemical and
biological systems, where complex structures (e.g., proteins or
polymeric carbon chains) are formed by joining smaller components
through local interactions. Here non-aggregative fusion abstracts the
joining of components not only at fixed attachment points, but also
through general, property-based interactions, modeled via color
compatibility in our framework. Studying the tree-width of these
structures is essential for enabling automated reasoning about their
properties. We consider the exploration of this area as future work.

\paragraph*{Related Work}
The notion of composition is central to substructural
logics~\cite{sep-logic-substructural}. One of the foremost such logics
is Bunched Implications (BI), whose first definition of semantics is
based on partially ordered monoids (i.e., the multiplicative
connective is interpreted as the multiplication in the
monoid)~\cite{PYM2004257}. In particular, the monoidal semantics of BI
(and many other follow-up logics) do not assume the composition to be
aggregative. The advent of the more popular semantics of BI based on
finite partial functions, called \emph{heaps}, has made aggregative
composition popular among the users of Separation
Logic~\cite{Ishtiaq00bias,Reynolds02}. We list below several
substructural logics where composition is not aggregative.

Docherty and Pym developped Intuitionistic Layered Graph Logic
(ILGL), a substructural logic tailored to reasoning about graph
structures with a fixed, non-commutative and non-associative notion
of layering~\cite{journals/lmcs/DochertyP18}.
%
Calcagno et al.~\cite{conf/popl/CalcagnoGZ05} introduce
Context Logic as a framework for local reasoning about structured
data, emphasizing compositionality through structural connectives
that describe data and context separately rather than flattening
them into aggregates. In~\cite{conf/popl/CalcagnoGZ07}, they
formalize these connectives as modal operators and demonstrate
that such non-aggregative reasoning is essential for expressing
weakest preconditions and verifying updates.
%
Cardelli et al.~\cite{Cardelli2002Spatial} present a spatial
logic for reasoning about graphs that, like our work, emphasizes
local, non-aggregative composition via structural connectives such
as spatial conjunction.


\section{Definitions}
\label{sec:definitions}

Given integers $i$ and $j$, we write $\interv{i}{j}$ for the set
$\set{i,i+1,\ldots,j}$, assumed to be empty if $i>j$. For a set $A$,
we denote by $\pow{A}$ its powerset. By writing $B \finsubseteq A$ we
mean that $B$ is a finite subset of $A$.  The cardinality of a finite
(multi)set $A$ is written $\cardof{A}$. By writing $A = A_1 \uplus
A_2$ we mean that $A_1$ and $A_2$ partition $A$, i.e., $A = A_1 \cup
A_2$ and $A_1 \cap A_2 = \emptyset$. The $n$-times Cartesian product
of $A$ with itself is denoted $A^n$ and the set of possibly empty
(resp. nonempty) sequences of elements from $A$ by $A^*$
(resp. $A^+$). 
Multisets are denoted as $\mset{a,b,\ldots}$, $\sqcup$
and $\sqcap$ denote the operations of multiset union and intersection,
respectively. The multi-powerset (i.e., the set of multisets) of $A$
is written $\mpow{A}$.



\subsection{Relational Structures}
\label{subsec:structures}


Let $\relations$ be a finite alphabet of \emph{relation symbols}
$\arel \in \relations$, of arities $\arityof{\arel} \geq 1$, and
$\constants$ be a countably infinite set of \emph{constants} $\acst
\in \constants$ of arity zero. As usual, relation symbols of arity
$1$, $2$ and $3$ are called unary, binary and ternary, respectively.
A $\csts$-\emph{structure}, for some finite set
$\csts\finsubseteq\constants$ of constants, is a pair
$\astruc=(\univ_\astruc,\struc_\astruc)$, where $\univ_\astruc$ is a
finite set called \emph{universe} and $\struc_\astruc$ is an
\emph{interpretation} that maps each relation symbol
$\arel\in\relations$ to a subset of $\univ^{\arityof{\arel}}_\astruc$
and each constant $\acst\in\csts$ to an element of
$\univ_\astruc$. The \emph{sort} of the structure $\astruc$ is the set
$\csts$. Two structures are \emph{disjoint} iff their universes are
disjoint and \emph{isomorphic} iff they are defined over the same
alphabet and differ only by a renaming of their elements\footnote{See,
e.g., \cite[Section A3]{DBLP:books/daglib/0082516} for a formal
definition of isomorphism between structures.}.  For a given sort
$\csts\finsubseteq\constants$, we denote by $\strucof{\rels}{\csts}$
the set of $\csts$-structures.

We define the \emph{composition} of two relational structures as the
component-wise union of disjoint isomorphic copies of the structures
followed by joining the elements that interpret the common constants.
This is the same as the \emph{gluing} operation defined by
Courcelle~\cite[Definition 2.1]{CourcelleVII}, that we recall below,
for self-completeness:


\begin{definition}\label{def:composition}
  Let $\astruc\in\strucof{\rels}{\csts}$ be a structure and $\sim~
  \subseteq \univ_\astruc \times \univ_\astruc$ be an equivalence
  relation, where $[u]_{\sim}$ denotes the equivalence class of
  $u\in\univ_\astruc$.  The \emph{quotient} of $\astruc$ with respect
  to $\sim$ is the $\csts$-structure $\astruc_{/\sim}$ defined
  as follows:
  \begin{align*}
  \univ_{\astruc_{/\sim}} \isdef & \set{[u]_\sim \mid u \in \univ_\astruc} \\
  \struc_{\astruc_{/\sim}}(\arel) \isdef & \set{([u_1]_{\sim}, \ldots, [u_{\arityof{\arel}}]_{\sim}) \mid
    (u_1, \ldots, u_{\arityof{\arel}}) \in \struc_\astruc(\arel)} \text{, for each } \arel\in\relations \\
  \struc_{\astruc_{/\sim}}(\acst) \isdef & [\struc_\astruc(\acst)]_\sim \text{, for each } \acst\in\csts
  \end{align*}
   Let $\astruc_i=(\univ_i,\struc_i)$ be disjoint
   $\csts_i$-structures, for $i=1,2$, and $\approx~ \subseteq (\univ_1
   \uplus \univ_2) \times (\univ_1 \uplus \univ_2)$ be the least
   equivalence relation such that $\struc_1(\acst) \approx
   \struc_2(\acst)$, for all $\acst\in\csts_1\cap\csts_2$. The composition of $\astruc_1$ with $\astruc_2$ is the
   $\csts_1\cup\csts_2$-structure $\astruc_1 \comp \astruc_2 = (\univ, \struc)$, where: \begin{align*}
     \univ \isdef & ~\set{[u]_\approx \mid u \in \univ_1 \uplus \univ_2} \\
     \struc (\arel) \isdef & ~\set{([u_1]_{\approx}, \ldots, [u_{\arityof{\arel}}]_{\approx}) \mid
       (u_1, \ldots, u_{\arityof{\arel}}) \in \struc_1(\arel) \uplus \struc_2(\arel)} \text{, for each } \arel\in\relations \\
     \struc(\acst) \isdef & ~[\struc_i(\acst)]_\approx \text{, if } \acst\in\csts_i \text{, for each } \acst\in\csts_1\cup\csts_2
     ~~\left(\text{i.e., } [\struc_1(\acst)]_\approx=[\struc_2(\acst)]_\approx \text{ if } \acst\in\csts_1\cap\csts_2\right)
   \end{align*}
\end{definition}
We remark that the composition of $\emptyset$-structures $\astruc_1$
and $\astruc_2$ is the same as their disjoint union, denoted
$\astruc_1 \uplus \astruc_2$.

The composition of structures is \emph{aggregative}, meaning that it
keeps both structures separate except for the interpretation of the
common constants. In the following, we define a non-aggregative
\emph{fusion} operation that matches also some of the elements which
are not interpretations of common constants.  In contrast to the
deterministic composition, the equivalence relation that matches
elements in the fusion operation is chosen nondeterministically.

Before formalizing the notion of non-aggregative fusion, we introduce
a generic mechanism for controlling which pairs of elements are
allowed to join.  We assume a designated set of unary relation symbols
$\colors \subseteq \relations$. The sets of relation symbols $\acolor
\in \pow{\colors}$ are called \emph{colors}.  Given some
$\csts$-structure $\astruc=(\univ_\astruc,\struc_\astruc)$, we denote
by $\colorof{\astruc}{}(u) = \{ \arel \in \colors \mid u \in
\struc_\astruc(\arel) \}$ the color of each element $u \in
\univ_\astruc$. Note that the empty set is a color.

Back to the definition of non-aggregative fusion, we use colors to
prevent joining elements labeled with non-disjoint colors. This is
captured by the following notion of compatibility:
\begin{definition}\label{def:compatible}
  Let $\astruc=(\univ,\struc)$ be a $\csts$-structure. A relation
  $\sim~ \subseteq \univ\times\univ$ is \emph{compatible} with
  $\astruc$ if and only if $\colorof{\astruc}{}(u_1) \cap
  \colorof{\astruc}{}(u_2) = \emptyset$, for each pair $u_1 \sim u_2$.
\end{definition}

We now define the non-aggregative fusion operation. The operation is
non-deterministic, i.e., returns a (possibly empty) set of structures.
For reasons of simplicity, the fusion operation is only defined for
structures of sort $\emptyset$, i.e., for structures that do not
interpret any constants. This restriction can be lifted at the expense
of complexifying the definition below, by considering fusion in which
the interpretation of common constants in both structures must always
be joined, as in Definition \ref{def:composition}.

Given disjoint sets $A$ and $B$, a relation $\sim~ \subseteq A \times
B$ is an \emph{$A$-$B$ matching} iff $\set{a,b} \cap \set{a',b'} =
\emptyset$, for all distinct pairs $(a,b),(a',b') \in ~\sim$.  The
least equivalence relation that contains $\sim$ is denoted
$\equiv_\sim \subseteq (A \uplus B) \times (A \uplus B)$. We say that
an equivalence relation $\equiv_\sim$ is \emph{$k$-generated} iff $\sim$ is a
matching consisting of $k$ pairs.

\begin{definition}\label{def:fusion}
  Let $\astruc_i = (\univ_i,\struc_i)$, for $i=1,2$, be two disjoint
  $\emptyset$-structures. The \emph{fusion} of $\astruc_1$ and
  $\astruc_2$ is the following set of $\emptyset$-structures:
  \begin{align*}
    \fuse{}{\astruc_1}{\astruc_2} \isdef
    \{(\astruc_1\comp\astruc_2)_{/{\equiv_\sim}} \mid & ~\sim \text{ non-empty } \univ_{1}\text{-}\univ_{2} \text{ matching compatible with } \astruc_1\comp\astruc_2\}
  \end{align*}
  Let $\strucs$ be a set of $\emptyset$-structures. The closure of
  $\strucs$ under fusion is the least set $\fclose{}{\strucs}$ such
  that $\strucs \cup \set{\fuse{}{\astruc_1}{\astruc_2} \mid
    \astruc_1,\astruc_2 \in \fclose{}{\strucs}}\subseteq
  \fclose{}{\strucs}$.
\end{definition}
Note that $\fuse{}{\astruc_1}{\astruc_2}=\emptyset$ iff
$\colorof{\astruc_1}{}(u_1) \cap \colorof{\astruc_1}{}(u_2) \neq
\emptyset$, for all pairs $(u_1,u_2) \in \univ_1 \times \univ_2$. The
problems considered in the rest of this paper concern the uses of
fusion, in addition to the composition and the unary operations on
structures introduced next.

\subsection{An Algebra of Structures}
\label{subsec:algebra}

We recall the definitions of sorted terms and
algebras~\cite[Definition 1.1]{CourcelleI}. Let $\sorts$ be a
countably infinite set of \emph{sorts} and let $\fsignature$ be a
countably infinite set of \emph{function symbols}, where the set
$\fsignature$ is called a \emph{signature}.  Each $f \in \fsignature$
has an associated tuple of argument sorts $\argof{f}$ and a value sort
$\sortof{f}$. The arity of $f$, denoted $\arityof{f}$, is the length
of $\argof{f}$. Moreover, each variable has a sort. A
$\fsignature$-term $t[x_1,\ldots,x_n]$ is built as usual from function
symbols and variables $x_1,\ldots,x_n$ of matching sorts. A ground
term is a term without variables.  A trivial term consists of a single
variable.  A term $t'$ is a subterm of $t$ iff there exists a term
$u[x]$ such that $t=u[t']$, where $u[t']$ denotes the replacement of
$x$ by $t'$ in $u$. The sort of a term $t$, denoted $\sortof{t}$ is
the value sort of the top-most symbol, i.e., either the value sort
$\sortof{f}$ of the top-most function symbol $f$, in case of a
non-trivial term $t$, or the sort $\sortof{x}$ of the variable $x$, in
case of a trivial term $x$. A position $p$ in a term $t$ is a node of
the tree that uniquely represents $t$, in the usual way (see
\cite{comon:hal-03367725} for a formal definition). $\termsof{\mathcal{F}}$
denotes the set of ground terms having function symbols taken from a
finite set $\mathcal{F} \finsubseteq \fsignature$.


An $\fsignature$-\emph{algebra} $\algof{A} =
(\set{\universeOf{A}_s}_{s\in\sorts},\set{f^\algof{A}}_{f \in
  \fsignature})$ consists of domains $\universeOf{A}_s$ of each sort
$s \in \sorts$ and interprets the function symbols $f \in \fsignature$
as functions $f^\algof{A} : \universeOf{A}_{s_1} \times \ldots \times
\universeOf{A}_{s_n} \rightarrow \universeOf{A}_{\sortof{f}}$, where
$\argof{f}=(s_1,\ldots,s_n)$. By the domain of $\algof{A}$ we
understand the set $\universeOf{A} = \bigcup_{s \in \sorts}
\universeOf{A}_s$. We denote by $t^\algof{A}$ the interpretation of an
$\fsignature$-term $t$ in $\algof{A}$, i.e., the function obtained by
replacing each function symbol that occurs in $t$ by its
interpretation. In particular, $t^\algof{A}$ is an element of the
domain of $\algof{A}$ if $t$ is ground.

\begin{figure}[t!]
  \vspace*{-.5\baselineskip}
  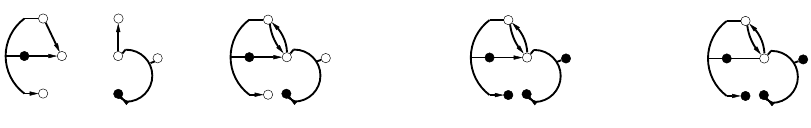
  \caption{HR operations on structures over the alphabet
    $\set{a,b,c}$, where $\arityof{a}=3$ and
    $\arityof{b}=\arityof{c}=2$. The order of vertices attached to an
    edge is indicated by an arrow pointing to the last vertex. }
  \label{fig:graphs}
  \vspace*{-.5\baselineskip}
\end{figure}

We define an algebra of structures, called $\algof{HR}$, with sorts
$\sorts_\algof{HR} \isdef \set{\csts \mid
  \csts\finsubseteq\constants}$, where each universe
$\universeOf{HR}_\csts$ is the set of $\csts$-structures. The
signature $\fsignature_\algof{HR}$ consists of: \begin{itemize}
\item constant symbols $\arel(\slabs_1, \ldots,
  \slabs_{\arityof{\arel}})$, for $\arel \in \relations$ and $\slabs_i
  \finsubseteq \constants$ such that either $\slabs_i = \slabs_j$ or
  $\slabs_i \cap \slabs_j = \emptyset$ for all $1 \le i < j \le
  \arityof{\arel}$, interpreted as $\arel(\slabs_1, \ldots,
  \slabs_{\arityof{\arel}})^\algof{HR} \isdef (\univ,\struc)$, where:
  \begin{align*}
    \univ \isdef & ~\set{u_1,\ldots,u_{\arityof{\arel}}}
    \text{ for some (possibly equal) elements } u_1,\ldots,u_{\arityof{\arel}} \\[-1mm]
    & \hspace*{2.2cm} \text{ such that } u_i = u_j \iff \slabs_i = \slabs_j \text{, for all } 1 \leq i<j \leq \arityof{\arel} \\
    \struc(\arel) \isdef & ~\set{(u_1,\ldots,u_{\arityof{\arel}})}
    \text{ and }
    \struc(\arel') \isdef \emptyset \text{, for all } \arel' \in \relations \setminus \set{\arel}
    \\
    \struc(\acst) \isdef & ~u_i \text{, for all } \acst \in \slabs_i \text{ and } 1 \leq i \leq \arityof{\arel}
  \end{align*}
\item binary function symbols $\pop{\csts}{\csts'}$, for $\csts,\csts'
  \finsubseteq\constants$, interpreted by the composition operation
  $\comp$ from \autoref{def:composition} applied to structures of
  sorts $\csts$ and $\csts'$, respectively, see Figure
  \ref{fig:graphs} (a).
\item unary function symbols $\rename{\alpha}{\csts}$, for all
  $\csts,\csts' \finsubseteq\constants$ and surjective function
  $\alpha : \csts \rightarrow \csts'$, interpreted as the operations
  $\rename{\alpha}{\csts}^\algof{HR}: \universeOf{HR}_\csts
  \rightarrow \universeOf{HR}_{\csts'}$ where, for each
  $\csts$-structure $\astruc$, the output $\astruc' =
  \rename{\alpha}{\csts}(\astruc)$ is defined below:
  \begin{align*}
    \univ_{\astruc'} \isdef &~ \univ_\astruc
    \hspace*{5mm}
    \struc_{\astruc'}(\arel) \isdef \struc_\astruc(\arel)
    \hspace*{5mm}
    \struc_{\astruc'}(\alpha(\acst)) \isdef \struc_\astruc(\acst)
    \text{, for all } \arel\in\relations \text{ and } \acst\in\csts\setminus\csts'
  \end{align*}
\item unary function symbols $\forget{\csts'}{\csts}$, for $\csts
  \finsubseteq\constants$ and $\csts' \subseteq \csts$, interpreted as
  the operations $\forget{\csts'}{\csts}^\algof{HR}:
  \universeOf{HR}_\csts \rightarrow
  \universeOf{HR}_{\csts\setminus\csts'}$ where, for each
  $\csts$-structure $\astruc$, the output $\astruc' =
  \forget{\csts'}{\csts}^\algof{HR}(\astruc)$ is defined below:
  \begin{align*}
    \univ_{\astruc'} \isdef &~ \univ_\astruc
    \hspace*{5mm}
    \struc_{\astruc'}(\arel) \isdef \struc_\astruc(\arel)
    \hspace*{5mm}
    \struc_{\astruc'}(\acst) \isdef \struc_\astruc(\acst)
    \text{, for all } \arel\in\relations \text{ and } \acst\in\csts
  \end{align*}
\end{itemize}
To ease the notation, we omit the sorts of the arguments from the
$\algof{HR}$ function symbols $\fsignature_\algof{HR}$ when they are
understood from the context.

\begin{example}
  The two leftmost graphs in Figure \ref{fig:graphs} (a) are the
  values of the following terms, respectively (singleton sets are denoted by their elements, to avoid clutter):
  \begin{align*}
    t_1 \isdef & ~\rename{\acst_3 \rightarrow \acst_1}{}\left(\forget{\acst_1}{}\left(a(\acst_0,\acst_1,\acst_2)
    \pop{}{} c(\acst_1,\acst_3)\right) \pop{}{} b(\acst_0,\acst_3)\right) \\
    t_2 \isdef & ~ b(\acst_0,\acst_1) \pop{}{} \forget{\acst_2}{}\left(a(\acst_1,\acst_3,\acst_2)\right)
  \end{align*}
  The graph in Figure \ref{fig:graphs} (c) is the value of the term
  $t_0 \isdef \rename{\acst_0 \leftrightarrow
    \acst_1}{}\left(\forget{\acst_2}{}(t_1 \pop{}{} t_2)\right)$.
\end{example}

We comment on the relationship to the algebra of relational structures
introduced in ~\cite[Definition 2.3]{CourcelleVII}, as a
generalization of both the hyperedge-replacement (\hr) algebra of
hypergraphs and the vertex-replacement (\vr) algebra of binary graphs,
to relational structures. As a matter of fact, we do not use this
algebra. Instead, our algebra of relational structures algebra
$\algof{HR}$ follows the standard \hr{} algebra on
hypergraphs~\cite{CourcelleI}. This relationship can be easily
understood by viewing relational structures as hypergraphs in the
adjacency encoding, i.e., each tuple $(u_1, \ldots,
u_{\arityof{\arel}})$ from the interpretation of a relation symbol
$\arel$ corresponds to a $\arel$-labeled hyperedge attached to the
vertices $u_1, \ldots, u_{\arityof{\arel}}$ in this order. For reasons
of space, we omit further details.

We now introduce the subalgebras of $\algof{HR}$ that define the
tree-width parameter of a structure. For this, we assume some
enumeration of the constants $\constants =
\set{\acst_0,\acst_1,\ldots}$ and denote by
$\fsignature_{\algof{HR}^{\le k}}$ the subset of
$\fsignature_\algof{HR}$ consisting of the function symbols whose
argument and value sorts are all contained in
$\set{\acst_0,\ldots,\acst_k}$.

\begin{definition}\label{def:tree-width}
  The \emph{tree-width} of a structure $\astruc$, denoted
  $\twof{\astruc}$, is the minimal integer $k\geq0$ for which there
  exists a ground term $t \in \termsof{\fsignature_{\algof{HR}^{\le
        k}}}$ such that $t^{\algof{HR}}=\astruc$. A set $\strucs$ of
  structures has \emph{bounded tree-width} if and only if the set
  $\set{\twof{\astruc} \mid \astruc \in \strucs}$ is finite.
\end{definition}
In particular, for each tree-width bounded set $\strucs$, there exists
a set $\mathcal{T}$ of ground terms and a finite set $\csts
\finsubseteq \constants$ of constants such that each term $t \in
\mathcal{T}$ uses only constants from $\csts$ and $\strucs =
\set{t^\algof{HR} \mid t \in \mathcal{T}}$.

This algebraic definition of tree-width of a relational structure is
analogous to the definition of the tree-width of a hypergraph using a
subalgebra of \hr{} defined by a restriction of sorts to finite sets
of vertex labels (see, e.g.,~\cite[Proposition
  1.19]{courcelle_engelfriet_2012} for a proof of equivalence between
the graph-theoretic and algebraic definitions of tree-width for
hypergraphs). Moreover, the tree-width of a structure can be
equivalently defined in terms of the tree-width of its Gaifman-graph:


\begin{definition}\label{def:gaifman}
   Let $\astruc$ be a $\csts$-structure.  The \emph{Gaifman graph} of
   $\astruc$ is the simple undirected graph $\gaifmanof{\astruc} =
   (\vertof{},\edgeof{})$, where $\vertof{} \isdef \univ_\astruc$ and
   $\edgeof{} \isdef \set{\set{u_i,u_j} \mid (u_1, \ldots,
     u_{\arityof{\arel}}) \in \struc_\astruc(\arel), 1 \leq i \neq j
     \leq \arityof{\arel}, \arel\in\relations}$.
\end{definition}
It is known that the tree-width of a structure equals the tree-width
of its Gaifman graph~\cite[Proposition
  11.27]{DBLP:series/txtcs/FlumG06}.

We recall below two standard notions of graph theory. Given binary
graphs $\graph$ and $H$, we say that $H$ is a \emph{minor} of $\graph$
iff $H$ is obtained from a subgraph of $\graph$ by \emph{edge
contractions}, where the contraction of a binary edge $e \in
\edgeof{\graph}$ attached to vertices $u$ and $v$ means deleting $e$
and joining $u$ and $v$ into a single vertex $x$ (the edges attached
to $x$ are the ones attached to either $u$ or $v$). It is well-known
that the tree-width of each minor of a graph $\graph$ is bounded by
the tree-width of $\graph$.

A $n \times m$-\emph{grid} is a binary graph whose vertices can be
labeled with pairs $(i,j) \in \interv{1}{n} \times \interv{1}{m}$ such
that there is an edge between $(i,j)$ and $(i',j')$ iff either $i <
n$, $i'=i+1$ and $j'=j$ or $j < m$, $i'=i$ and $j'=j+1$.  It is
well-known\footnote{This can be shown by using the characterisation of
tree-width in terms of the cops and robber game~\cite{SEYMOUR199322}.}
that each $n \times n$-grid has tree-width $n$. By bluring the
distinction between isomorphic grids, we obtain the following:

\begin{proposition}\label{prop:grids-tw}
  A set of structures whose Gaifman graphs contain infinitely many
  non-isomorphic square grids has unbounded tree-width.
\end{proposition}


\subsection{Context-Free Sets of Structures}
\label{subsec:context-free}

Context-free sets are usually defined as languages of grammars, i.e.,
finite sets of inductive rules written using nonterminals and function
symbols from a given signature. To simplify some of the upcoming
proofs, we use here an equivalent definition of contex-free sets based
on recognisable sets of ground terms, defined using tree
automata~\cite{comon:hal-03367725}. For self-containment reasons, we
briefly introduce context-free grammars and discuss their equivalence
with tree automata in Section \ref{subsec:colabs} (see the statement
of Theorem \ref{thm:ft}).

Let $\sign \finsubseteq \fsignature$ be a finite signature of function
symbols. A \emph{tree automaton} over $\sign$ is a tuple $\mathcal{A}
= (Q,F,\arrow{}{})$, where $Q$ is a finite set of states, $F \subseteq
Q$ is a set of accepting states and $\arrow{}{}$ is a set of
transition rules of the form $(q_1,\ldots,q_{\arityof{f}}) \arrow{f}{}
q$, where $q_1,\ldots,q_{\arityof{f}},q \in Q$ and $f\in\sign$. A run
$\pi$ of $\mathcal{A}$ over a ground term $t \in \termsof{\sign}$ maps
each position $p$ within $t$ to a state $q=\pi(p)$ if the automaton
has a rule $(\pi(p_1), \ldots, \pi(p_{\arityof{f}})) \arrow{f}{} q$,
where $p_1, \ldots, p_{\arityof{f}}$ are the positions of the children
of $p$ in $t$. A ground term $t$ is accepted by $\mathcal{A}$ iff
$\mathcal{A}$ has a run that labels the root of $t$ with an accepting
state. The language of $\mathcal{A}$, denoted $\lang{\mathcal{A}}$,
is the set of ground terms accepted by $\mathcal{A}$. A set $T
\subseteq \termsof{\sign}$ is \emph{recognisable} iff it is the
language of a tree automaton over the finite signature $\sign$.

\begin{definition}
  A set $\strucs$ of structures is \emph{context-free} if and only
  if there exists a recognisable set of ground terms $T \subseteq
  \termsof{\sign}$, over a finite signature $\sign \subseteq
  \fsignature_\algof{HR}$, such that $\strucs = \set{t^\algof{HR}
    \mid t \in T}$.
\end{definition}
Note that a context-free set of structures has finitely many sorts,
because the signature of terms used to describe the set is finite. For
this reason, any context-free set of structures has bounded
tree-width. On the other hand, there are bounded tree-width sets which
are not context-free, for instance the set of linear structures over
the alphabet $\set{a,b,c}$ of binary relations of the form
$a^nb^nc^n$, for all $n\geq1$.

\section{The Closure of Context-Free Sets under Fusion}

This section is concerned with the statement and proof of the main
result of the paper (Theorem \ref{thm:main}). For simplicity, we
assume that $\strucs$ is a set of \emph{connected} structures, where
connectivity of a structure $\astruc$ means that there exists an
undirected path in $\gaifmanof{\astruc}$ between each pair of elements
$u_1,u_2 \in \univ_\astruc$. We note that, because $\strucs$ contains
only connected structures, each structure from its closure
$\fclose{}{\strucs}$ is necessarily connected.

The assumption of $\strucs$ being a context-free set of connected
structures loses no generality, because it is possible, from a tree
automaton $\mathcal{A}$ such that $\strucs=\lang{A}^\algof{HR}$, to
build a tree automaton $\mathcal{B}$ such that $\lang{B}^\algof{HR}$
is the set of connected substructures of a structure in
$\strucs$. Intuitively, $\mathcal{B}$ is obtained from $\mathcal{A}$
be labeling each state $q$ with finite information concerning the
existence of a path between each pair of constant symbols in each
structure that is the value of a ground term recognized by $q$ in
$\mathcal{A}$. This construction can be used to generalize the
statement of Theorem \ref{thm:main} below from connected to arbitrary
structures. We will detail this construction in an extended version
of the present article.

\begin{theorem}\label{thm:main}
  Let $\strucs$ be a context-free set of connected
  $\emptyset$-structures.
  \begin{enumerate}[1.]
  \item\label{it1:thm:main} $\fclose{}{\strucs}$ has bounded
    tree-width if and only if $\fclose{}{\strucs}$ is
    context-free.
  \item\label{it2:thm:main} It is decidable whether
    $\fclose{}{\strucs}$ has bounded tree-width.
  \end{enumerate}
\end{theorem}
An obvious consequence of this theorem is the decidability of the
problem: \emph{given a context-free set $\strucs$, is $\fclose{}{\strucs}$
context-free?}

We give an overview of the proof before going into technical
details. The core idea is the equivalence between the (\ref{eq:condA})
tree-width boundedness of the closure $\fclose{}{\strucs}$ of a
context-free set $\strucs$ and (\ref{eq:condB}) the non-existence of
two structures $\astruc_1,\astruc_2 \in \fclose{}{\strucs}$ having
each at least three elements each $u_i,v_i,w_i \in \univ_{\astruc_i}$,
labeled with disjoint colors $\acolor_i$, for $i=1,2$. This
equivalence is established via a third, more technical, condition
about the disjointness relations between the colors that may occur in
a structure from $\fclose{}{\strucs}$.  The latter implies that the
matching relation of each fusion of two structures from
$\fclose{}{\strucs}$ is generated by at most two pairs of elements
with compatible colors.

The equivalence between (\ref{eq:condA}) and (\ref{eq:condB}) is used
to prove both points of Theorem \ref{thm:main}. For point
(\ref{it1:thm:main}) suppose, for a contradiction, that
(\ref{eq:condB}) does not hold. Then, $\astruc_1$ and $\astruc_2$ can
be composed by joining their $u_i$, $v_i$ or $w_i$ elements, for
$i=1,2$, respectively, as in Figure \ref{fig:grids}. Consequently, the
set of Gaifman graphs corresponding to the structures in
$\fclose{}{\strucs}$ contains an infinite set of grid minors, thus
$\fclose{}{\strucs}$ has unbounded tree-width (Proposition
\ref{prop:grids-tw}). Else, if (\ref{eq:condB}) holds (i.e., such
structures cannot be found), we prove that the matching relation
considered in the fusion of any two structures is generated by either
one or two pairs of elements. In each of these cases, by adding a
finite number of constants to the signature of the
$\fsignature_{\algof{HR}}$-terms from the language of $\mathcal{A}$,
we can build a tree automaton $\mathcal{A}^*$ such that
$\lang{\mathcal{A}^*}^\algof{HR}=\fclose{}{\strucs}$, thus
taking care of point (\ref{it1:thm:main}) of the theorem.

To prove point (\ref{it2:thm:main}), we rely on the equivalence of
(\ref{eq:condA}) and (\ref{eq:condB}) and show that (\ref{eq:condB})
is decidable. This is done by arguing that the existence of two
structures with the above property is equivalent to the existence of
two multiset abstractions of structures
$\mset{\acolor_1,\acolor_1,\acolor_1}$,
$\mset{\acolor_2,\acolor_2,\acolor_2}$ in the domain of multisets of
colors having multiplicity at most three. These multiset abstractions
of colors can be effectively computed by a finite fixpoint iteration
over the rules of the tree automaton that recognizes the set of ground
terms which evaluates to the elements of $\strucs$.

\subsection{Color Multisets}

In the following, let $\strucs$ be a context-free set of
$\emptyset$-structures.  We denote the set of colors by
$\allcols\isdef\pow{\colors}$, where $\colors$ is a fixed finite set
of unary relation symbols. First, we define an abstraction of
structures as finite multisets of colors:

\begin{definition}\label{def:multiset-color-abstraction}
  The \emph{multiset color abstraction} $\mcolabs{\astruc} \in
  \mpow{\allcols}$ of a structure $\astruc$ is $\mcolabs{\astruc}
  \isdef \mset{ \colorof{}{\astruc}(u) \mid u \in
    \univ_\astruc}$.
  For an integer $k\geq0$, the \emph{$k$-multiset
    color abstraction} $\kmcolabs{k}{\astruc} \subseteq
  \mpow{\allcols}$ is $\kmcolabs{k}{\astruc} \isdef \set{ M \subseteq
    \mcolabs{\astruc} \mid \cardof{M} \le k}$.
    These abstractions are
  lifted to sets of structures as sets of multisets $\mcolabs{\strucs}
  \isdef \set{ \mcolabs{\astruc} \mid \astruc \in {\strucs}}$ and
  $\kmcolabs{k}{\strucs} \isdef \bigcup_{\astruc \in \strucs}
  \kmcolabs{k}{\astruc}$.
\end{definition}
Note that $\mcolabs{\astruc}$ is a multiset, whereas
$\kmcolabs{k}{\astruc}$ is a set of multisets. When lifted to sets of
structures, both $\mcolabs{\strucs}$ and $\kmcolabs{k}{\strucs}$ are
sets of multisets.

\begin{figure}[t!]
  \vspace*{-.5\baselineskip}
  \centerline{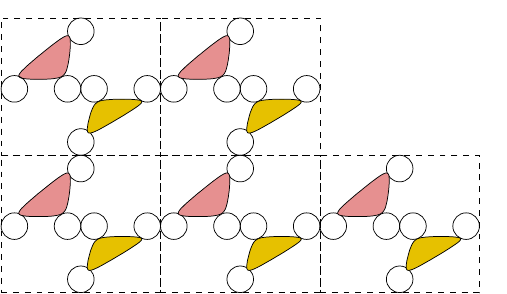}
  \caption{Building structures whose Gaifman graphs have infinitely large grid minors}
  \label{fig:grids}
  \vspace*{-.5\baselineskip}
\end{figure}

The core of the proof of Theorem \ref{thm:main} is the equivalence
between the following two conditions stated formally below: 
\begin{align}
  \twof{\fclose{}{\strucs}} \leq & ~k \text{, for some } k \geq 1 \tag{A}\label{eq:condA} \\
  \mset{\acolor_1,\acolor_1,\acolor_1},\mset{\acolor_2,\acolor_2,\acolor_2}
  \in \kmcolabs{3}{(\fclose{}{\strucs})} \Rightarrow & ~\acolor_1 \cap
  \acolor_2 \neq \emptyset \text{, for all } \acolor_1,\acolor_2\in\allcols \tag{B}\label{eq:condB}
\end{align}
Note that condition (\ref{eq:condA}) is more general than the premiss
of Theorem \ref{thm:main}. We prove the (\ref{eq:condA}) $\Rightarrow$
(\ref{eq:condB}) direction below.

\begin{lemma}\label{lemma:grid-minor}
  If $\strucs$ has bounded tree-width, then (\ref{eq:condA}) implies (\ref{eq:condB}).
\end{lemma}
\begin{proof}
  By contradiction, assume that there exist
  $\mset{\acolor_1,\acolor_1,\acolor_1},\mset{\acolor_2,\acolor_2,\acolor_2}
  \in \kmcolabs{3}{(\fclose{}{\strucs})}$ such that $\acolor_1 \cap
  \acolor_2 = \emptyset$. Then, there exist structures
  $\astruc_1,\astruc_2 \in \fclose{}{\strucs}$ such that
  $\mset{\acolor_1,\acolor_1,\acolor_1} \in \mcolabs{\astruc}_1$ and
  $\mset{\acolor_2,\acolor_2,\acolor_2} \in \mcolabs{\astruc}_2$. We
  shall use $\astruc_1$ and $\astruc_2$ to build infinitely many
  structures whose Gaifman graphs have arbitrarily large square grid
  minors, as illustrated in Figure \ref{fig:grids}.

  First, construct the structure ${\astruc_{12}} \in
  \fclose{}{\strucs}$ by fusing one pair $(u_1,u_2)$, having colors
  $\acolor_1$ and $\acolor_2$, respectively. Let $v_1$ and $w_1$
  (resp. $v_2$ and $w_2$) be the remaining distinct elements of
  $\astruc_{12}$, having color $\acolor_1$ (resp. $\acolor_2$) from
  $\astruc_1$ and $\astruc_2$, respectively. For an arbitrarily large
  integer $n\geq1$, consider $n\times n$ disjoint copies
  $(\astruc_{12}^{i,j})_{i,j=1,n}$ of $\astruc_{12}$. Let
  $\approx^{1,j}$ be the equivalence relation generated by
  $\set{(v_1^{1,j}, v_2^{1,j-1})}$ and $\approx^{i,1}$ be generated by
  $\set{(w_2^{i,1},w_1^{i-1,1})}$, $\approx^{i,j}$ be generated by
  $\set{(v_1^{i,j}, v_2^{i,j-1}), (w_2^{i,j},w_1^{i-1,j})}$, for all
  $i,j=2,n$. Second, construct the grid-like connected structure
  $X^{n,n} \in \fclose{}{\strucs}$:
  \[
    X^{n,n} = (...( ... ((\astruc_{12}^{1,1}
    \comp \astruc^{1,2}_{12})_{/\approx^{1,2}} \comp
    \astruc^{2,1}_{12})_{/\approx^{2,1}} \comp...  \comp
    \astruc^{i,j}_{12})_{/\approx^{i,j}} \comp ...  \comp
    \astruc^{n,n}_{12})_{/\approx^{n,n}}
  \]
  where structures $\astruc^{i,j}_{12}$ are added to the fusion in the
  increasing order of $i+j$. We can show that $\gaifmanof{X^{n,n}}$
  has an $n \times n$ square grid minor. Finally, as $n$ can be taken
  arbitrarily large, we conclude that $\fclose{}{\strucs}$ does not
  have bounded tree-width, which contradicts (\ref{eq:condA}).
\end{proof}

\subsection{Color Schemes}

\begin{figure}[t!]
  \vspace*{-.5\baselineskip}
  \centerline{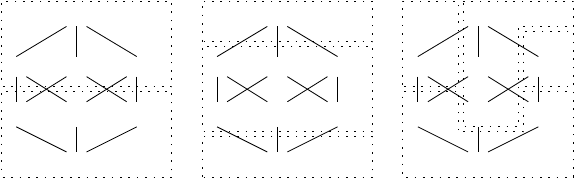}
  \caption{Examples of RGB color schemes}
  \label{fig:rgb-color-schemes}
  \vspace*{-.5\baselineskip}
\end{figure}

For the proof of the (\ref{eq:condA}) ``$\Leftarrow$''
(\ref{eq:condB}) direction, we first organize the set of colors using
the \emph{RGB color schemes} defined below:

\begin{definition}\label{def:rgb-color-scheme}
  A partition $(\redcols,\greencols,\bluecols)$ of $\allcols$ is an
  \emph{RGB color scheme} if and only if:
  \begin{enumerate}
  \item $\acolor_1 \cap \acolor_2 \not= \emptyset$, for all
    $\acolor_1,\acolor_2 \in \bluecols$,
  \item $\acolor_1 \cap \acolor_2 \not= \emptyset$, for all $\acolor_1
    \in \greencols$ and all $\acolor_2 \in \bluecols$,
  \item for all $\acolor_1 \in \redcols$ there exists $\acolor_2 \in
    \bluecols$ such that $\acolor_1 \cap \acolor_2 = \emptyset$.
  \end{enumerate}
\end{definition}
Note that an RGB color scheme is fully specified by the set
$\bluecols$. Indeed, any color not in $\bluecols$ is unambiguously
placed within $\redcols$ or $\greencols$, depending on whether or not
it is disjoint from some color in $\bluecols$. In particular, if
$\bluecols=\emptyset$ then $\redcols=\emptyset$ and
$\greencols=\allcols$. For example, \autoref{fig:rgb-color-schemes}
shows several RGB color schemes for the set
$\colors=\set{\aarel,\abrel,\acrel}$ of unary relation symbols.

Because a fusion operation only joins element with disjoint colors,
blue elements can only be joined with red elements, green elements can
be joined with green or red elements, whereas red elements can be
joined with elements of any other color. We define below what is meant
for a set of structures to conform to an RGB color scheme:

\begin{definition}\label{def:conforming-structures}
  A set $\strucs$ of structures \emph{conforms to
  $(\redcols,\greencols,\bluecols)$} if and only if: \begin{enumerate}
  \item\label{it1:def:conforming-structures} for all structures
    $\astruc\in\strucs$, if $\colorof{}{\astruc}(u)\in\redcols$, for some
    element $u \in \univ_\astruc$, then
    $\colorof{}{\astruc}(u')\in\bluecols$, for all other elements $u' \in
    \univ_\astruc \setminus \set{u}$, and
  \item\label{it2:def:conforming-structures} $\mcolabs{\astruc} \cap
    \greencols \subseteq \mset{ \acolor, \acolor \mid \acolor \in
      \greencols}$, for all structures $\astruc \in \fclose{}{\strucs}$.
  \end{enumerate}
\end{definition}
In other words, $\strucs$ conforms to a given color scheme if each
structure from $\strucs$ has either a single red and the rest blue, or
at most occurrences of the same green color and the rest blue
elements. Moreover, the number of occurrences of a green color must
not exceed two, for each structure obtained by taking fusions of some
structures in $\strucs$. This observation justifies the following
notion of \emph{type} of a structure:

\begin{definition}\label{def:structure-type}
  A structure $\astruc$ is of \emph{type} $\redtype$ if it has exactly
  one red element and the rest blue, $\greentype$ if it has at least
  one green element and the rest blue and $\bluetype$ if it has only
  blue elements.
\end{definition}
Note that there can be structures of neither $\redtype$, $\greentype$
or $\bluetype$ type, but these are the only types of interest, as
justified by the following:

\begin{lemma}\label{lemma:it1:rgb-conformance}
  Let $\strucs$ be a set of structures conforming to an RGB color
  scheme. Then, each structure $\astruc \in \fclose{}{\strucs}$ is of
  type either $\redtype$, $\greentype$ or $\bluetype$.
\end{lemma}
\begin{proof}
  By induction on the construction
  of $\astruc \in \fclose{}{\strucs}$ from one or more
  structures from $\strucs$. \autoref{table:extfusion-types}
  summarizes the possible types of
  $\fuse{}{\astruc_1}{\astruc_2}$ on structures $\astruc_1$
  and $\astruc_2$ of types $\redtype$, $\greentype$ or $\bluetype$,
  respectively.
  \begin{table}[t!]
    \centering
    \vspace*{-.5\baselineskip}
    \begin{tabular}{|c|c|c|c|}
      \hline
      $\fuse{}{\astruc_1}{\astruc_2}$ &
      $\astruc_2$ of $\redtype$ type &
      $\astruc_2$ of $\greentype$ type &
      $\astruc_2$ of $\bluetype$ type \\
      \hline 
      $\astruc_1$ of $\redtype$ type &
      $\redtype, \greentype, \bluetype$ &
      $\greentype, \bluetype$ &
  $\bluetype$ \\ 
      $\astruc_1$ of $\greentype$ type &
      $\greentype, \bluetype$ &
      $\greentype, \bluetype$ & $\emptyset$ \\ 
      $\astruc_1$ of $\bluetype$ type &
      $\bluetype$ & $\emptyset$ & $\emptyset$ \\ 
      \hline
    \end{tabular}
    \caption{The types of structures obtained by fusion, where $\emptyset$
      means that the result of the fusion is the empty
      set.}\label{table:extfusion-types}
    \vspace*{-.5\baselineskip}
  \end{table}
\end{proof}

The (\ref{eq:condB}) ``$\Rightarrow$'' (\ref{eq:condA}) direction will
be established via a third condition (\ref{eq:condC}), which is
conformance to an RGB color scheme defined by taking the $\bluecols$
set to be the colors occurring three times in some structure from
$\fclose{}{\strucs}$:

\begin{lemma}\label{lemma:btw-rgb-conformance}
  If (\ref{eq:condB}) holds then:
  \begin{align}
    \strucs \text{ conforms to }
    (\redcols,\greencols,\bluecols) \text{, where } \bluecols \isdef \{\acolor\in\allcols \mid
    \mset{\acolor,\acolor,\acolor} \in
    \kmcolabs{3}{(\fclose{}{\strucs})}\}\tag{C}\label{eq:condC}
  \end{align}
\end{lemma}
\begin{proof}
  We show that $\strucs$
  conforms to the $(\redcols,\greencols,\bluecols)$ RGB color scheme
  from the statement, by checking the two points of
  \autoref{def:conforming-structures}: \begin{itemize}
  \item (\ref{it1:def:conforming-structures}) Let $\astruc \in
    \strucs$ and prove that for any two colors
    $\acolor_1,\acolor_2\in\colors$, if $\mset{\acolor_1, \acolor_2}
    \subseteq \mcolabs{\astruc}$ and $\acolor_1 \in \redcols$ then
    $\acolor_2 \in \bluecols$. Since $\acolor_1 \in \redcols$, there
    must exists a color $\acolor_1' \in \bluecols$, such that
    $\acolor_1 \cap \acolor_1' = \emptyset$, by
    \autoref{def:rgb-color-scheme}. By the definition of $\bluecols$,
    this further implies $\mset{\acolor_1', \acolor_1',\acolor_1'} \in
    \kmcolabs{3}{(\fclose{}{\strucs})}$.  Henceforth, there exists a
    structure $\astruc' \in \fclose{}{\strucs}$ such that
    $\mset{\acolor_1', \acolor_1',\acolor_1'} \subseteq
    \mcolabs{\astruc'}$. We can now use $\astruc'$ and three disjoint
    copies of $\astruc$ to build a new structure $\astruc''$ by gluing
    progressively, each one of the three elements of color
    $\acolor_1'$ in $\astruc'$ to the element of color $\acolor_1$ of
    $\astruc$. Then, by construction, the structure $\astruc''$ will
    also contain three elements of color $\acolor_2$, one from each
    disjoint copy of $\astruc$.  Therefore, $\mset{\acolor_2,
      \acolor_2, \acolor_2} \in \mcolabs{\astruc''}$ and because
    $\astruc'' \in \fclose{}{\strucs}$ this implies $\mset{\acolor_2,
      \acolor_2, \acolor_2} \in \kmcolabs{3}{(\fclose{}{\strucs})}$
    and therefore $\acolor_2 \in \bluecols$.
  \item (\ref{it2:def:conforming-structures}) By contradiction, let
    $\astruc \in \fclose{}{\strucs}$ be such that $\mcolabs{\astruc}
    \sqcap \greencols \not\subseteq \mset{ \acolor, \acolor \mid
      \acolor \in \greencols}$.  Then there exists $\acolor' \in
    (\mcolabs{\astruc} \sqcap \greencols) \setminus \mset{ \acolor,
      \acolor \mid \acolor \in \greencols}$, i.e., $\acolor' \in
    \greencols$ and $\mset{\acolor',\acolor',\acolor'} \subseteq
    \mcolabs{\astruc}$. The latter implies
    $\mset{\acolor',\acolor',\acolor'} \in \kmcolabs{3}{\astruc}
    \subseteq \kmcolabs{3}{(\fclose{}{\strucs})}$.  But this implies
    $\acolor' \in \bluecols$ according to the definition of the RGB
    color scheme, contradicting $\acolor' \in \greencols$.
  \end{itemize}
\end{proof}
In the rest of this and the next subsections, we are concerned with
the proof of the following implication, that establishes the
equivalence of (\ref{eq:condA}) and (\ref{eq:condB}). As previously
mentioned, this direction of the proof uses the third condition
(\ref{eq:condC}) that is, conformance to the RGB color scheme from the
statement of Lemma \ref{lemma:btw-rgb-conformance}:

\begin{lemma}\label{lemma:btw-equivalence}
  If $\strucs$ has bounded tree-width and (\ref{eq:condC}) holds then
  (\ref{eq:condA}) holds.
\end{lemma}

\begin{figure}[t!]
  \vspace*{-.5\baselineskip}
  \centerline{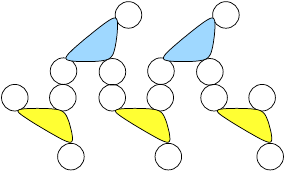}
  \caption{Fusion of $\greentype$ structures by
  $3$-generated matchings}
  \label{fig:green}
  \vspace*{-.5\baselineskip}
\end{figure}

The proof of the above lemma is split into two technical results
(Lemmas \ref{lemma:it2:rgb-conformance} and
\ref{lemma:it3:rgb-conformance}). The first (Lemma
\ref{lemma:it2:rgb-conformance}) involves reasoning about the number
of pairs of elements that are joined by the fusion operation in order
to obtain a structure from $\fclose{}{\strucs}$:

\begin{lemma}\label{lemma:it2:rgb-conformance}
  If $\strucs$ conforms to an RGB scheme
  $(\redcols,\greencols,\bluecols)$ and $\astruc = (\astruc_1 \uplus
  \astruc_2)_{/\approx}$ for some $\astruc_i = (\univ_i,\struc_i) \in
  \fclose{}{\strucs}$, for $i=1,2$, and some equivalence closure
  $\approx$ of a $\univ_{1}$-$\univ_{2}$ matching, then exactly one of
  the following holds: \begin{enumerate}
  \item\label{it21:lemma:rgb-conformance} $\approx$ is
    $1$-generated, or
  \item\label{it22:lemma:rgb-conformance} $\approx$ is $2$-generated,
    $\astruc$ is of type $\bluetype$, and either: \begin{enumerate}
    \item\label{it221:lemma:rgb-conformance} $\astruc_1$,
      $\astruc_2$ are both of type $\redtype$, or
    \item\label{it222:lemma:rgb-conformance} $\astruc_1$,
      $\astruc_2$ are both of type $\greentype$ and
      $\cardof{\mcolabs{\astruc_1} \sqcap \greencols} =
      \cardof{\mcolabs{\astruc_2} \sqcap \greencols} = 2$.      
    \end{enumerate}
    \end{enumerate}
\end{lemma}
\begin{proof}
  We distinguish two cases:
  \begin{itemize}
  \item $\astruc_1$ is of type $\redtype$: If $\astruc_2$ is of type
    $\bluetype$ or $\greentype$ then $\astruc_1$ and $\astruc_2$ can
    be fused only by equivalences $\approx$ generated by a single
    pair, that contains the element from the support of $\astruc_1$
    with color in $\redcols$, thus matching the case
    (\ref{it21:lemma:rgb-conformance}) from the statement. Else, if
    $\astruc_2$ is of type $\redtype$ then $\astruc_1$ and $\astruc_2$
    can be fused by equivalences generated by at most two pairs, each
    containing an element with color from $\redcols$, from either
    $\astruc_1$ or $\astruc_2$, thus matching the case
    \ref{it221:lemma:rgb-conformance} from the statement. In this
    latter case, $\astruc$ is of type $\bluetype$ because joining a
    red with a blue element always results in a blue element.
  \item $\astruc_1$, $\astruc_2$ are both of type $\greentype$: By
    contradiction, assume they can be fused by an equivalence
    $\approx$ generated by three pairs of elements
    $(u_{1i},u_{2i})_{i=1,2,3}$. Let
    $G_{1i}=\colorof{}{\astruc_1}(u_{1i})$, $G_{2i}
    =\colorof{}{\astruc_2}(u_{2i})$ be the colors from $\greencols$ of
    the matching elements in the two structures, for $i=1,2,3$.  Then,
    we can construct structures using $\astruc_1$ and $\astruc_2$
    where any of these colors repeat strictly more than twice,
    henceforth, contradicting the conformance property to the RGB
    color scheme. The principle of the construction is depicted in
    \autoref{fig:green}. Finally, note that the construction depicted
    in \autoref{fig:green} fuse actually only pairs of colors
    $(G_{1i},G_{2i})$ for $i=1,2$. Henceforth, the conformance
    property is also contradicted if $\astruc_1$ and $\astruc_2$ can
    be fused by a $2$-generated equivalence relation $\approx$, such
    that the support of either $\astruc_1$ or $\astruc_2$ contains
    more than three elements with colors in $\greencols$. By the same
    argument, it follows that $\astruc$ is of type $\bluetype$, if
    $\approx$ is generated by two pairs of green elements. 
  \end{itemize}
  \vspace*{-\baselineskip}
\end{proof}

\subsection{Fusions as Operations on Terms}
      
The second technical result required for the proof of Lemma
\ref{lemma:btw-equivalence} is a characterization of the $k$-generated
fusion, for $k=1,2$, via operations on witness
$\fsignature_\algof{HR}$-terms. We assume that $\strucs$ is a given
tree-width bounded set of structures that conforms to a fixed RGB
color scheme $(\redcols,\greencols,\bluecols)$. The goal is to prove
that $\twof{\fclose{}{\strucs}} \leq \twof{\strucs}+K$, for an integer
$K\geq0$.

\begin{figure}[t!]
  \vspace*{-.5\baselineskip}
  \centerline{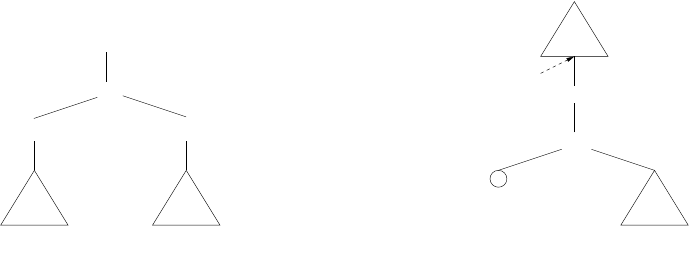}
  \caption{The
    $\joinop{t_1}{t_2}{\acst^{i_{1}}_{\acolor_{1}}}{\acst^{i_{2}}_{\acolor_2}}$
    (a) and $\appendop{t_1}{t_2}{n}{k}{\acst^i_\acolor}$ (b)
    operations}\label{fig:ops}
  \vspace*{-.5\baselineskip}
\end{figure}

Since $\strucs$ has bounded tree-width, there exists a set
$\mathcal{T}$ of terms such that $\strucs = \set{t^\algof{HR} \mid t
  \in \mathcal{T}}$ and a finite set $\csts$ of constants such that
each term $t \in \mathcal{T}$ uses only constants from $\csts$ and
assume w.l.o.g. $\csts$ to be the least such set. Moreover, consider
the following set of \emph{special} constants $\overline{\csts} \isdef
\set{\acst^i_\acolor \mid \acolor \in \allcols,~ 1 \leq i \leq 2}$,
with the following intuition. Recall that each element of a structure
$t^\algof{HR}$ is the common interpretation of all the constants
$\acst \in \slabs_i$ in the label $\arel(\slabs_1, \ldots,
\slabs_{\arityof{\arel}})$ of a leaf of $t$. By adding
$\acst^i_\acolor$ to the set $\slabs_i$, we mean that the color of
that element in the structure is $\acolor$. A constant is
\emph{visible} in a term if it is not in the scope of a $\rename{}{}$
or $\forget{}{}$ operation. Let $K \isdef \cardof{\overline{\csts}}$
and note that $\cardof{K} = 2\cdot\cardof{\allcols}$.


We shall prove that $\twof{\astruc} \leq \twof{\strucs} + K$ by
building, for any $\astruc \in \fclose{}{\strucs}$, a term $t$ that
uses only constants from $\csts \uplus \overline{\csts}$, such that
$\forget{\overline{\csts}}{}\left(t^\algof{HR}\right) = \astruc$. We
then note that $\astruc$ has tree-width at most $\cardof{\csts}+K$ and
obtain $\twof{\astruc} \leq \cardof{\csts} + K =
\twof{\strucs}+K$. Since the choice of $\astruc\in\fclose{}{\strucs}$
was arbitrary, this leads to $\twof{\fclose{}{\strucs}} \leq
\twof{\strucs}+K$.

In the following, we understand terms as trees whose nodes are labeled
by function symbols from $\fsignature_\algof{HR}$. For each node $n$
of a term $t$, we write $\nodelab{n}$ for its label. The children of
each node $n$ form an ordered sequence of length equal to the arity of
$\nodelab{n}$. In order to the build the witness terms for the
structures $\astruc \in \fclose{}{\strucs}$, we make use of the
following operations on terms $t$, $t_1$ and $t_2$ having constants in
$\csts \uplus \overline{\csts}$: \begin{enumerate}
\item $\labelop{t}{n}{k}{\acst^i_\acolor}$, where $n$ is a leaf of
  $t$, $\nodelab{n}=\arel(\csts_1,\ldots,\csts_{\arityof{\arel}})$ and
  $k \in \interv{1}{\arityof{\arel}}$: 
  Let $1 \leq k_1 < \ldots < k_\ell \leq \arityof{\arel}$ be the
  indices of the sets of constants from $\nodelab{n}$ that are equal
  to $\csts_k$ (see the definition of $\algof{HR}$ in Section
  \ref{subsec:algebra}). The operation changes the label of $n$ in $t$
  only, by replacing the set $\csts_{k_j}$ with $\csts_{k_j} \cup
  \set{\acst^i_\acolor}$, for each $1 \leq j \leq \ell$.
  %
\item $\joinop{t_1}{t_2}{\acst^{i_{1}}_{\acolor_{1}}}{\acst^{i_{2}}_{\acolor_2}}$:
  The result is the following term, for a nondeterministic choice of $j$, such that
  $\acst^{j}_{\acolor_{1} \uplus \acolor_{2}}$ is not visible in either $t_1$ or $t_2$ (the operation is undefined otherwise):
  \[
  \hspace*{-4mm}\forget{\mathcal{B}}{}(\rename{\acst^{i_{1}}_{\acolor_{1}} \rightarrow~ \acst^{j}_{\acolor_{1} \uplus \acolor_{2}}}{}(t_1) ~\pop{}{}
  \rename{\acst^{i_{2}}_{\acolor_{2}} \rightarrow~ \acst^{j}_{\acolor_{1} \uplus \acolor_{2}}}{}(t_2)),~ 
  \mathcal{B} \isdef \set{\acst^{j}_{\acolor_{1} \uplus \acolor_{2}} \mid \acolor_{1} \uplus \acolor_{2} \in \bluecols}
  \]
  We refer to Figure \ref{fig:ops} (a) for an illustration. In
  addition, we consider an overloaded version of
  $\joinop{t_1}{t_2}{\acst^{i_{11}}_{\acolor_{11}},\acst^{i_{12}}_{\acolor_{12}}}{\acst^{i_{21}}_{\acolor_{21}},\acst^{i_{22}}_{\acolor_{22}}}$
  that fuses the interpretation of $\acst^{i_{1j}}_{\acolor_{1j}}$
  with that of $\acst^{i_{2j}}_{\acolor_{2j}}$, for both $j=1,2$. This
  definition is similar to the one above, thus omitted for brevity.
\item $\appendop{t_1}{t_2}{n}{k}{\acst^i_\acolor}$, where $n$ is a
  leaf of $t_1$, $\nodelab{n} = \arel(\csts_1,\ldots,\csts_{\arityof{\arel}})$ and $k \in \interv{1}{\arityof{\arel}}$:
  the result is the term
  $t_1[n/\forget{\acst^i_\acolor}{}(\labelop{n}{n}{k}{\acst^i_\acolor}
    \pop{}{} t_2)]$, where $t[n/s]$ denotes the substitution of the
  leaf $n$ by the term $s$ in $t$. We refer to Figure \ref{fig:ops}
  (b) for an illustration.
\end{enumerate}
Then, Lemma \ref{lemma:btw-equivalence} is an immediate consequence of
the following lemma:

\begin{lemma}\label{lemma:it3:rgb-conformance}
  For each structure $\astruc \in \fclose{}{\strucs}$ there exists a
  $\fsignature_\algof{HR}$-term $t$ using only constants from $\csts
  \cup \overline{\csts}$, such that (i)
  $\forget{\overline{\csts}}{}\left(t^\algof{HR}\right) = \astruc$ and
  (ii) for each element $u \in \univ_\astruc$ such that
  $\acolor=\colorof{}{\astruc}(u)\in \redcols \uplus \greencols$ there
  exists a special constant $\acst^i_\acolor \in \overline{\csts}$
  such that $\struc_\astruc(\acst^i_\acolor) = u$.
\end{lemma}
\begin{proof}\emph{(sketch)}
  We build $t$ by induction of the derivation of $\astruc =
  (\univ,\struc) \in \fclose{}{\strucs}$. For the base case
  $\astruc\in\strucs$, let $t' \in \mathcal{T}$ be a term such that
  $\astruc = {t'}^\algof{HR}$. By repeating the
  $\labelop{t}{n}{k}{\acst^i_\acolor}$ operation, we add a special
  constant $\acst^i_\acolor$ to each leaf $n$ of $t$, on the
  appropriate position $1 \leq k \leq \arityof{\arel}$, where
  $\nodelab{n} = \arel(\csts_1,\ldots,\csts_{\arityof{\arel}})$, such
  that $\struc(\csts_k) = \set{u}$ and $\acolor \isdef
  \colorof{}{\astruc}(u) \in \redcols \uplus \greencols$. The choice
  of $1 \leq i \leq 2$ is nondeterministic. The result of applying
  these labeling operations to $t'$ is $t$. Then,
  $\forget{\overline{\csts}}{}\left(t^\algof{HR}\right) = \astruc$ and
  (ii) holds, by construction.
  For the inductive step, let $\astruc=(\astruc_1 \uplus
  \astruc_2)_{/\approx}$, where
  $\astruc_1,\astruc_2\in\fclose{}{\strucs}$ and $\approx$ is an
  equivalence relation that is generated by the set of pairs
  $\set{(u_{1i},u_{2i})}_{i\in I}$, where $I$ is either $\set{1}$ or
  $\set{1,2}$. Let $\acolor_{ji} \isdef \colorof{}{\astruc_j}(u_{ji})$
  for all $1 \leq j \leq 2$ and $i \in I$. By the inductive hypothesis,
  there exist terms $t_j$ and integers $1 \leq k_{ji} \leq 2$, such
  that $(\univ_j,\struc_j) \isdef t_j^\algof{HR}$ and
  $\forget{\overline{\csts}}{}\left(t_j^\algof{HR}\right) =
  \astruc_j$, for all $1 \leq j \leq 2$ and $i \in I$. \begin{enumerate}
  \item $I=\set{1}$, i.e., $\approx$ is $1$-generated. \begin{enumerate}
    \item\label{item11:rgb-conformance} $\acolor_{11},\acolor_{21} \in
      \redcols\uplus\greencols$: by the inductive hypothesis (ii),
      there exist $\acst^{i_1}_{\acolor_{11}},
      \acst^{i_2}_{\acolor_{21}} \in \overline{\csts}$ such that
      $u_{j1} = \struc_j(\acst^{i_j}_{\acolor_{j1}})$, for both $1
      \leq j \leq 2$. Suppose that $\acst^\ell_{\acolor_{11} \uplus
        \acolor_{21}}$ is visible in $t_1$ (visibility in $t_2$ is a
      symmetric case), for some $1 \leq \ell \leq 2$. Then
      $\acolor_{11}\uplus\acolor_{21}$ must belong to $\redcols \uplus
      \greencols$, by the inductive hypothesis (ii). If
      $\acst^{3-\ell}_{\acolor_{11} \uplus \acolor_{21}}$ is not
      visible in $t_2$, suppose first that
      $\acst^{3-\ell}_{\acolor_{11} \uplus \acolor_{21}}$ is visible
      in $t_1$. Then, $\acst^{i_1}_{\acolor_{11}}$,
      $\acst^\ell_{\acolor_{11} \uplus \acolor_{21}}$ and
      $\acst^{3-\ell}_{\acolor_{11} \uplus \acolor_{21}}$ are visible
      in $t_1$. By Lemma \ref{lemma:it1:rgb-conformance},
      $\acolor_{11},\acolor_{11} \uplus \acolor_{21}\in\greencols$ and
      $\acolor_{11}\uplus\acolor_{21}$ occurs $3$ times in $\astruc$,
      thus contradicting the definition of $\bluecols$. Hence
      $\acst^{3-\ell}_{\acolor_{1} \uplus \acolor_{2}}$ is not visible
      in either $t_1$ or $t_2$ and $t \isdef
      \joinop{t_1}{t_2}{\acst^{i_1}_{\acolor_{11}}}{\acst^{i_2}_{\acolor_{21}}}$
      is well-defined. Else, if $\acst^{3-\ell}_{\acolor_{11} \uplus
        \acolor_{21}}$ is visible in $t_2$, then
      $\acolor_{11}\uplus\acolor_{21}$ occurs $3$ times in $\astruc$,
      thus contradicting the definition of $\bluecols$.
    \item\label{item12:rgb-conformance} $\acolor_{11} \in \bluecols$
      and $\acolor_{21} \in \redcols$ ($\acolor_{11} \in \redcols$ and
      $\acolor_{21} \in \bluecols$ is a symmetric case): Let $n$ be
      the leaf of $t_1$ such that $\nodelab{n} = \arel(\csts_1,
      \ldots, \csts_{\arityof{\arel}})$ and $\csts_k$ be a set of
      constants that are (all) interpreted as $u_{11}$, for some $1
      \leq k \leq \arityof{\arel}$. Let $\acst^\ell_{\acolor_{21}}$ be
      the special constant such that $u_{21} =
      \struc_2(\acst^\ell_{\acolor_{21}})$, for some $1 \leq \ell \leq
      2$. We can assume w.l.o.g. that $\acst^\ell_{\acolor_{21}}$ is
      not visible in $n$. If this were not to be the case, then $n$
      must have been involved in a previous join of a term $t_3$ with
      another term $t'_1$, such that $t_1$ is the outcome of this
      join. In this case, we change the construction, by first joining
      $t_2$ with $t_3$, as in the previous case, then joining the
      result with $t'_1$,. Note that this is possible due of the
      associativity of the $\pop{}{}$ operation. Finally, we define $t
      \isdef \appendop{t_1}{t_2}{n}{k}{\acst^\ell_{\acolor_{21}}}$.
    \end{enumerate}
  \item $I=\set{1,2}$, i.e., $\approx$ is $2$-generated. By Lemma
    \ref{lemma:it2:rgb-conformance}, either one of the following
    holds: \begin{enumerate}
    \item\label{item21:rgb-conformance} $\astruc_1$ and $\astruc_2$
      are of type $\greentype$: let $\acst^{k_{ji}}_{\acolor_{ji}}$ be
      special constants such that $u_{ji} =
      \struc_j(\acst^{k_{ji}}_{\acolor_{ji}})$, for all $1 \leq i,j
      \leq 2$. Then, we define $t \isdef
      \joinop{t_1}{t_2}{(\acst^{k_{i}}_{\acolor_{1i}})_{i\in
          I}}{(\acst^{k_{i}}_{\acolor_{2i}})_{i\in I}}$ and check that
      the operation is well-defined, following a similar argument as
      in case (\ref{item11:rgb-conformance}).
    \item\label{item22:rgb-conformance} $\astruc_1$ and $\astruc_2$
      are of type $\redtype$: we can assume w.l.o.g. that $u_{ii}$ is
      the interpretation of a special constant
      $\acst^{j_{i}}_{\acolor_{ii}}$, for some $1 \leq j_i \leq 2$,
      where $\acolor_{ii} \in \redcols$, for both $i=1,2$. Let $n_1$
      be the leaf of $t_1$ such that $\nodelab{n_1} = \arel(\csts_1,
      \ldots, \csts_{\arityof{\arel}})$ and $u_{11}$ is the
      interpretation of (all) constants from $\csts_{k_1}$, for some
      $1 \leq k_1 \leq \arityof{\arel}$. Analogously, we consider
      $n_2$ to be the leaf of $t_2$ and $k_2$ the position of the
      constants from its label, that are interpreted as
      $u_{22}$. Under similar assumptions as in the case
      (\ref{item21:rgb-conformance}), ensuring that the result is well
      defined, we let $t \isdef
      \forget{\acst^{j_1}_{\acolor_{11}}}{}(\appendop{t_1}{\labelop{t_2}{n_2}{k_2}{\acst^{j_1}_{\acolor_{11}}}}{n_1}{k_1}{\acst^{j_2}_{\acolor_{22}}})$.
      The only difference with the previous case is that the indices
      $j_1$ and $j_2$ must be different to avoid name clashes, hence
      we require $2$ special constants $\acst^1_\acolor$ and
      $\acst^2_\acolor$, for each color $\acolor\in\redcols$.
      \vspace*{-\baselineskip}
    \end{enumerate}
  \end{enumerate}
\end{proof}

\subsection{Tree-width Bounded Fusion-closed Sets are Context-free}

This subsection completes the proof of the first point of Theorem
\ref{thm:main}. The final ingredient is the following lemma, whose
proof relies on the lifting of the construction that simulates the
$1$- or $2$-generated fusion of structures from terms to tree automata
recognising sets of terms:

\begin{lemmaE}\label{lemma:fusion-context-free}
  Let $\strucs$ be a context-free set of structures conforming to some
  RGB color scheme. Then, the set $\fclose{}{\strucs}$ is
  context-free.
\end{lemmaE}
\begin{proofE}
  Let $(\redcols,\greencols,\bluecols)$ be the RGB color scheme from
  the hypothesis and $\mathcal{A}=(Q,F,\arrow{}{A})$ be a tree
  automaton that recognizes a set of terms over a finite signature
  $\mathcal{F} \subseteq \fsignature_\algof{HR}$, such that $\strucs =
  \set{t^\algof{HR} \mid t \in \lang{\mathcal{A}}}$. Let $\csts$ be
  the finite set of constants that occur in some term from
  $\termsof{\mathcal{F}}$. By pre-processing $\mathcal{A}$, we assume
  w.l.o.g. that each state $q \in Q$ carries (i.e., is annotated with)
  the following information, that holds in each run $\pi$ of
  $\mathcal{A}$ over a term $t\in\termsof{\mathcal{F}}$, such that
  $q$ is the state labeling the root of $t$: \begin{enumerate}[i.]
  \item\label{it1:pre-processing} $t$ consists of a single leaf node
    labeled $\arel(\slabs_1,\ldots,\slabs_k)$ such that the (common)
    interpretation $u_\ell$ of the constants in each set $\slabs_\ell
    \subseteq \csts$ in the structure $t^\algof{HR}$ has color
    $\acolor_\ell$, i.e., $\colorof{}{t^\algof{HR}}(u_\ell) =
    \acolor_\ell$. This fact is denoted as
    $\colorof{}{q}(\slabs_\ell)=\acolor_\ell$, by abuse of
    notation. This information can be inferred by tracking bottom-up
    the visibility and identity (through $\rename{\slabs}{}$ and
    $\forget{\alpha}{}$ operations) of each constant $\acst \in
    \slabs_\ell$, while guessing its final color. The automaton has a
    run over $t$ only if all such guesses are correct, otherwise it
    gets stuck. Moreover, we denote by $\leafof{q}\in\set{\top,\bot}$ the
    boolean information whether $q$ labels a leaf subterm in each run.
  \item\label{it2:pre-processing} $t \in \lang{\mathcal{A}}$ (i.e., $q
    \in F$) and $t^\algof{HR}$ is a structure of type $\redtype$,
    $\greentype$ or $\bluetype$. This information can be inferred by
    counting the number of red and green elements up to or more than
    $1$, based on the information on colors provided at the previous
    point. We denote by $\typeof{q} \in
    \set{\redtype,\greentype,\bluetype}$ this information.
  \end{enumerate}
  Note that this pre-processing step involves splitting the
  transitions of $\mathcal{A}$ over multiple copies of the same
  original state having different annotations.

  We build a tree automaton $\mathcal{A}^*$ over the finite
  signature: \begin{align*} \overline{\mathcal{F}} \isdef &
    ~\set{\arel(\slabs'_1,\ldots,\slabs'_k) \mid
      \arel(\slabs_1,\ldots,\slabs_k) \in \mathcal{F},~ \slabs_i
      \subseteq \slabs'_i \subseteq \slabs_i \uplus \overline{\csts}
      \text{, for all } 1 \leq i \leq k} \\
    & ~\cup \set{f \in \mathcal{F} \mid f = \forget{\slabs'}{\slabs''},~
      \slabs',\slabs''\subseteq\csts\cup\overline{\csts}} \\
    & ~\cup \set{f \in \mathcal{F} \mid f = \rename{\alpha}{\slabs'},~ \alpha :
      \csts\cup\overline{\csts} \rightarrow \csts\cup\overline{\csts},~
      \slabs'\subseteq\csts\cup\overline{\csts}}
  \end{align*}
  such that $\fclose{}{\strucs} = \set{t^\algof{HR} \mid t \in
    \lang{\mathcal{A}^*}}$. The construction of
  $\mathcal{A}^*$ follows the structure of the inductive
  proof of Lemma \ref{lemma:it3:rgb-conformance}.

  The states and transitions of $\mathcal{A}^*$ contain the
  states and some of the transitions of $\mathcal{A}$,
  respectively. In addition, $\mathcal{A}^*$ has states of
  the form $(m,P,q)$, where: \begin{itemize}
  \item $m \in \set{\amode, \amode_1, \amode'_1, \amode_2, \amode_3}$
    indicates in which \emph{mode} the automaton $\mathcal{A}^*$ is
    currently: $\amode$ is the default mode, in which the special
    constants are propagated upwards, $\amode_1$ and $\amode'_1$ are
    used to build the gadget of cases (\ref{item11:rgb-conformance})
    and (\ref{item21:rgb-conformance}), $\amode_2$ the one of case
    (\ref{item12:rgb-conformance}) and $\amode_3$ is used for case
    (\ref{item22:rgb-conformance}) of the proof of Lemma
    \ref{lemma:it3:rgb-conformance}. The mode indicator allows the
    $\mathcal{A}^*$ to finish building these intermediate terms
    correctly, before re-entering the default mode.
  \item $P \subseteq \overline{\csts}$ is a set of special constants,
    that are visible from the current position in the term,
  \item $q \in Q \cup \set{\bot}$ is either the currently tracked
    state of $\mathcal{A}$, or $\bot$. Intuitively, $q \in Q$
    indicates that $\mathcal{A}^*$ is currently reading a
    term from $\lang{\mathcal{A}}$, whereas $q=\bot$ means that
    $\mathcal{A}^*$ has finished reading the terms from
    $\lang{\mathcal{A}}$.
    %
    %
  \end{itemize}

  \begin{figure}[h!]
    \centerline{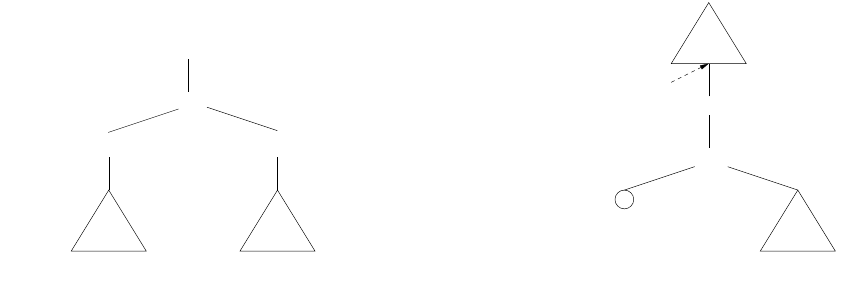}
    \caption{The implementation of the
      $\joinop{t_1}{t_2}{\acst^{i_{1}}_{\acolor_{1}}}{\acst^{i_{2}}_{\acolor_2}}$
      (a) and $\appendop{t_1}{t_2}{n}{k}{\acst^i_\acolor}$ (b)
      operations in $\mathcal{A}^*$, where the states of
      $\mathcal{A}^*$ are depicted in red.}
    \label{fig:auto}
  \end{figure}

  In addition to transitions inherited from $\mathcal{A}$, the
  automaton $\mathcal{A}^*$ has the following transitions,
  defined according to the cases of the inductive proof of Lemma
  \ref{lemma:it3:rgb-conformance}: \begin{itemize}
  \item For the base case, $\mathcal{A}^*$ has the following
    transitions: \[\begin{array}{rll}
    \arrow{\arel(\slabs'_1,\ldots,\slabs'_{\arityof{\arel}})}{} & (\amode,P,q,\typeof{q})
    & \text{for each transition } \arrow{\arel(\slabs_1,\ldots,\slabs_{\arityof{\arel}})}{} q \text{ of } \mathcal{A}, \\
    && \text{where } \acolor_k \isdef \colorof{}{q}(\slabs_k) \\
    && P \isdef \set{\acst^{i_k}_{\acolor_k} \mid \acolor_k \not\in\bluecols,~ 1 \leq k \leq \arityof{\arel}} \\
    && \slabs'_k \isdef \left\{\begin{array}{ll}
    \slabs_k \uplus \set{\acst^{i_k}_{\acolor_k}} & \text{if } \acolor_k \in \redcols \uplus \greencols \\
    \slabs_k & \text{if } \acolor_k \in \bluecols
    \end{array}\right. \\
    && \text{such that } i_k \in \left\{\begin{array}{ll}
    \interv{1}{2\arityof{\arel}} & \text{if } \acolor_k \in \redcols \\
    \set{1,2} & \text{if } \acolor_k \in \greencols
    \end{array}\right. \\
    && \text{for all } 1 \leq k \leq \arityof{\arel}
    \\\\[-2mm]
    (\amode,P,q) \arrow{f}{} & (\amode,P,q')
    & \text{for each transition } q \arrow{f}{} q' \text{ of } \mathcal{A} \\
    && \text{and each } P \subseteq \overline{\csts}
    \\\\[-2mm]
    ((\amode,P,q_1), q_2) \arrow{\pop{}{}}{} & (\amode,P,q)
    & \text{for each transition } (q_1,q_2) \arrow{\pop{}{}}{} q \text{ of } \mathcal{A}, \\
    && \text{each } P \subseteq \overline{\csts} \text{ and } q_1 \in Q 
    \\\\[-2mm]
    (q_1,(\amode,P,q_2)) \arrow{\pop{}{}}{} & (\amode,P,q) & \text{idem}
    \\\\[-2mm]
    ((\amode,P_1,q_1),(\amode,P_2,q_2)) \arrow{\pop{}{}}{} & (\amode, P_1 \uplus P_2, q, \hastype)
    & \text{for each transition } (q_1,q_2) \arrow{\pop{}{}}{} q \text{ of } \mathcal{A}
    \end{array}\]
    Moreover, the transitions
    $\arrow{\arel(\slabs_1,\ldots,\slabs_{\arityof{\arel}})}{} q$ such that
    $\colorof{}{q}(\slabs_\ell) \in \redcols \uplus \greencols$ for
    some $1 \leq \ell \leq \arityof{\arel}$ (i.e., inherited from $\mathcal{A}$) are
    removed, because red and green elements must be always initially
    marked by special constants. These constants are then propagated
    upwards during each run of $\mathcal{A}^*$.
  \item The cases (\ref{item11:rgb-conformance}) and
    (\ref{item21:rgb-conformance}) of the proof of Lemma
    \ref{lemma:it3:rgb-conformance} are concerned with joining two
    structures via a $1$- or $2$-generated equivalence relation,
    respectively. We give the construction of $\mathcal{A}^*$ in the
    first case, the second being similar. Consider that these
    structures are evaluations of $\mathcal{F}$-terms $t_1$ and $t_2$
    that have been labeled by $\mathcal{A}^*$ with states
    $(\amode,P_1,q_1)$ and $(\amode,P_2,q_2)$, respectively. Then, the
    new term is the result of the nondeterministic operation
    $\joinop{t_1}{t_2}{\acst^{k_{j}}_{\acolor_{1}}}{\acst^{k_{j}}_{\acolor_{2}}}$,
    implemented in $\mathcal{A}^*$ by the following transitions, for
    each $j \in \set{1,2}$:
    \[\begin{array}{rl}
    (\amode,P_k,q_k) \arrow{\rename{[\acst^{i_{k}}_{\acolor_{k}} \rightarrow~ \acst^{j}_{\acolor_{1} \uplus \acolor_{2}}]}{}}{} & (\amode_1,P'_k,q_k) \text{, where} \\
    & P'_k \isdef (P_k \setminus \set{\acst^{i_{k}}_{\acolor_{k}}}) \cup \set{\acst^{j}_{\acolor_{1} \uplus \acolor_{2}}} \text{, for both } k=1,2
    \\\\[-2mm]
    ((\amode_1, P'_1, q_1), (\amode_1, P'_2, q_2)) \arrow{\pop{}{}}{} & (\amode'_1, P', \bot) \text{, where } P' \isdef P'_1 \cup P'_2
    \\\\[-2mm]
    (\amode'_1, P',\bot) \arrow{\forget{\mathcal{B}}{}}{} & (\amode, P' \setminus \mathcal{B},\bot) \text{, where }
    \mathcal{B} \isdef \set{\acst^{j}_{\acolor_{1} \uplus \acolor_{2}} \mid \acolor_{1} \uplus \acolor_{2} \in \bluecols}
    \end{array}\]
    We refer to Figure \ref{fig:auto} (a) for an illustration of this construction.
  \item The case (\ref{item12:rgb-conformance}) of the proof of
    Lemma \ref{lemma:it3:rgb-conformance} is concerned with joining a red
    element of $t_2^\algof{HR}$ with a blue element of
    $t_1^\algof{HR}$ via a special constant $\acst^k_{\acolor_{2}}$,
    that is immediately forgotten after the join, because the
    resulting element is blue. The new term is the result of an
    operation $\appendop{t_1}{t_2}{n}{k}{\acst^k_{\acolor_{2}}}$,
    where $n$ is a leaf of $t_1$ having label $\nodelab{n} =
    \arel(\slabs_1,\ldots,\slabs_{\arityof{\arel}})$ and $k \in
    \interv{1}{\arityof{\arel}}$ is the position of that leaf
    corresponding to the blue element of $t_1^\algof{HR}$ involved in
    the fusion. This operation is implemented in
    $\mathcal{A}^*$ by the following transitions:
    \[\begin{array}{rl}
    \arrow{\arel(\slabs'_1,\ldots,\slabs'_{\arityof{\arel}})}{} & (\amode,P_0 \cup \set{\acst^k_{\acolor_{2}}}, q_0) \text{, where} \\
    & \slabs'_\ell \isdef \left\{\begin{array}{ll}
    \slabs_\ell \cup \set{\acst^k_{\acolor_{2}}} & \text{if } \slabs_\ell = \slabs_k \\
    \slabs_\ell & \text{otherwise}
    \end{array}\right. \text{, for all } 1 \leq \ell \leq \arityof{\arel} \\
    & \arrow{\arel(\slabs_1,\ldots,\slabs_{\arityof{\arel}})}{} (\amode,P_0,q_0) \text{ was added to } \mathcal{A}^* \text{ by the base case}
    \\\\[-2mm]
    ((\amode,P_0 \cup \set{\acst^k_{\acolor_{2}}},q_0),(\amode,P_2,q_2)) \arrow{\pop{}{}}{} & (\amode_2,P',\bot) \text{, where} \\
    & \aleaf(q_0) = \top,~ \acst^k_{\acolor_2} \not\in P_0,~  \acst^k_{\acolor_2} \in P_2 \\
    & P' \isdef P_0 \cup P_2
    \\\\[-2mm]
    (\amode_2,P',\bot) \arrow{\forget{\set{\acst^k_{\acolor_2}}}{}}{} & (\amode,P'',\bot) \text{, where }
    P'' \isdef P' \setminus \set{\acst^k_{\acolor_2}}
    \end{array}\]
    Finally, we add a new transition
    $L[(\amode,P_0,q_0)/(\amode,P'',\bot)] \arrow{f}{} (\amode,P,q)$
    for each transition $L \arrow{f}{} (\amode,P,q)$ introduced at the
    construction for the base case, where
    $L[(\amode,P_0,q_0)/(\amode,P'',\bot)]$ denotes the substitution
    of the state $(\amode,P_0,q_0)$ with the state $(\amode,P'',\bot)$
    on the left-hand side of the transition. We refer to Figure
    \ref{fig:auto} (b) for an illustration of this construction.
  \item The case (\ref{item22:rgb-conformance}) of the proof of Lemma
    \ref{lemma:it3:rgb-conformance} is concerned with joining a red
    element, that interprets the constant $\acst^{k_i}_{\acolor_{ii}}$
    from the structure $\astruc_i$, with a blue element from the
    structure $\astruc_{3-i}$, for both $i=1,2$. The new term is the
    result of an operation
    $\forget{\acst^{k_1}_{\acolor_{11}}}{}(\appendop{t_1}{\labelop{t_2}{n_2}{k_2}{\acst^{k_1}_{\acolor_{11}}}}{n_1}{k_1}{\acst^{k_2}_{\acolor_{22}}})$,
    where $n_i$ is the leaf of $t_i$ and $k_i$ is the position within
    that leaf of the blue element of $\astruc_i$ involved in the
    $2$-generated fusion. This operation
    $\appendop{t_1}{\labelop{t_2}{n_2}{k_2}{\acst^{k_1}_{\acolor_{11}}}}{n_1}{k_1}{\acst^{k_2}_{\acolor_{22}}}$
    is implemented in $\mathcal{A}^*$ by a similar construction to the
    one at the previous point, the only difference being the extra
    transitions needed to implement
    $\labelop{t_2}{n_2}{k_2}{\acst^{k_1}_{\acolor_{11}}}$ and
    propagate $\acst^{k_1}_{\acolor_{11}}$ to the root of $t_2$. The
    final $\forget{\acst^{k_1}_{\acolor_{11}}}{}(\ldots)$ is
    implemented as in the previous case (the last transition).
  \end{itemize}
  The automaton $\mathcal{A}^*$ has a single final state
  $q_f$ that is reached from each state $(m,P,q)$ such that $q$ is
  either a final state of $\mathcal{A}$ or $\bot$, via an operation
  that forgets all the special constants from $\overline{\csts}$. We
  sketch the proof of $\fclose{}{\strucs} = \set{t^\algof{HR} \mid t
    \in \lang{\mathcal{A}^*}}$ below:

  \vspace*{.5\baselineskip}\noindent ``$\subseteq$'' For each
  structure $\astruc \in \fclose{}{\strucs}$ there exists a
  $\mathcal{F}$-term $t$ such that $t^\algof{HR} =
  \overline{\astruc}$, by the argument used in the proof of Lemma
  \ref{lemma:it3:rgb-conformance}. This term is labeled by
  $\mathcal{A}^*$ with a state of the form $(m,P,q)$, by the
  construction of $\mathcal{A}^*$. Then $q$ is a final state
  of $\mathcal{A}$ if $\astruc\in\strucs$ and $q=\bot$ otherwise. In
  both cases, $\mathcal{A}^*$ has a transition to its final
  state, in order to accept the term
  $\forget{\overline{\csts}}{}(t)$. Since
  $\astruc=\forget{\overline{\csts}}{}^\algof{HR}(t^\algof{HR})$, we
  obtain $\astruc \in \set{t^\algof{HR} \mid t \in
    \lang{\mathcal{A}^*}}$.

  \vspace*{.5\baselineskip}\noindent ``$\supseteq$'' Let
  $t\in\lang{\mathcal{A}^*}$ be a
  $\overline{\mathcal{F}}$-term and $\astruc \isdef t^\algof{HR}$ be
  the structure obtained by evaluating $t$. Let $(m,P,q)
  \arrow{\forget{\overline{\csts}}{}}{} q_f$ be the last transition of
  $\mathcal{A}^*$ on some accepting run over $t =
  \forget{\overline{\csts}}{}(\overline{t})$, for some
  $\overline{\mathcal{F}}$-term $\overline{t}$. Then, we have
  ${\overline{t}}^\algof{HR} = \overline{\astruc}$. If $q \in F$ is a
  final state of $\mathcal{A}$, by the construction of
  $\mathcal{A}^*$, we have $t' \in \lang{\mathcal{A}}$, where
  $t'$ is the term obtained by removing the special constants from the
  labels of the leaves of $\overline{t}$. Hence, $t^\algof{HR} =
  {t'}^\algof{HR} \in \strucs \subseteq \fclose{}{\strucs}$.

  Otherwise, $q=\bot$ and $\overline{t}$ is a term built inductively
  from terms in $\lang{\mathcal{A}}$ annotated with special constants,
  as in the proof of Lemma \ref{lemma:it3:rgb-conformance}. Then
  $\overline{\astruc'} = \overline{t}^\algof{HR}$, where $\astruc'$ is
  the structure agreeing with $\overline{t}^\algof{HR}$, except for
  the interpretation of the special constants, where
  $\struc_{\astruc'}$ is undefined. By the choice of $\overline{t}$,
  $\astruc'$ is obtained from structures in $\strucs$ by inductively
  applying a $1$- or $2$-generated fusion, i.e., $\astruc' \in
  \fclose{}{\strucs}$. Since $t =
  \forget{\overline{\csts}}{}(\overline{t})$, we have $\astruc' =
  t^\algof{HR} = \astruc$, thus $\astruc \in \fclose{}{\strucs}$.
\end{proofE}

\noindent\textbf{Proof of Theorem \ref{thm:main} (\ref{it1:thm:main})}
``$\Rightarrow$'' Since $\strucs$ is a context-free set, it has
bounded tree-width. If $\fclose{}{\strucs}$ has bounded tree-width,
then $\strucs$ conforms to an RGB color scheme, by the combined
results of Lemmas \ref{lemma:grid-minor} and
\ref{lemma:btw-rgb-conformance}. Because $\strucs$ is context-free, we
obtain that $\fclose{}{\strucs}$ is context-free, by Lemma
\ref{lemma:fusion-context-free}. ``$\Leftarrow$'' Because each
context-free set has bounded tree-width.

\subsection{Color Abstractions}
\label{subsec:colabs}

This section is concerned with the proof of the second point of
Theorem~\ref{thm:main}. Let $\strucs$ be a context-free set of
structures given by a tree automaton $\mathcal{A}$ over a finite
signature $\mathcal{F} \subseteq \fsignature_\algof{HR}$. In the rest
of this section, $\strucs$, $\mathcal{F}$ and $\mathcal{A}$ are
considered to be fixed.

Note that the equivalence between (\ref{eq:condA}) and
(\ref{eq:condB}) follows from (\ref{eq:condA}) $\Rightarrow$
(\ref{eq:condB}) (Lemma \ref{lemma:grid-minor}), (\ref{eq:condB})
$\Rightarrow$ (\ref{eq:condC}) (Lemma \ref{lemma:btw-rgb-conformance})
and (\ref{eq:condC}) $\Rightarrow$ (\ref{eq:condA}) (Lemma
\ref{lemma:btw-equivalence}). Hence, it is sufficient to establish the
decidability of the condition (\ref{eq:condB}) for $\strucs$. To this
end, we compute the $k$-multiset abstraction
$\kmcolabs{k}{(\fclose{}{\strucs})}$, for an arbitrary given integer
$k\geq1$ (note that checking (\ref{eq:condB}) requires $k=3$).  First,
we reduce the computation of $\kmcolabs{k}{(\fclose{}{\strucs})}$ to
that of $\kmcolabs{k}{\strucs}$. Second, we sketch the argument behind
the effective computability of $\kmcolabs{k}{\strucs}$.

The following lemma shows that, because we are interested only in
$k$-multisets color abstractions, we can restrict fusion to
$1$-generated equivalence relations, while preserving the $k$-multiset
color abstraction. We denote by $\fclose{1}{\strucs}$ the set of
structures obtained by taking the closure of $\strucs$ only with
respect to fusions induced by $1$-generated matchings.

\begin{lemmaE}\label{lemma:reach-fusion-one}
  $\kmcolabs{k}{(\fclose{}{\strucs})} =
  \kmcolabs{k}{(\fclose{1}{\strucs})}$ for any set $\strucs$ of
  structures and integer $k\geq1$.
\end{lemmaE}
\begin{proofE}
  ``$\kmcolabs{k}{(\fclose{1}{\strucs})} \subseteq
  \kmcolabs{k}{(\fclose{}{\strucs})}$'' This direction follows
  directly from $\fclose{1}{\strucs} \subseteq
  \fclose{}{\strucs}$.

  ``$\kmcolabs{k}{(\fclose{}{\strucs})}
  \subseteq \kmcolabs{k}{(\fclose{1}{\strucs})}$'' We prove
  the stronger property:
\[
  \forall \astruc \in \fclose{}{\strucs}.~
  \exists \astruc' \in \fclose{1}{\strucs}.~
  \mcolabs{\astruc} \subseteq \mcolabs{\astruc'}
\]
By induction on the derivation of
$\astruc\in\fclose{}{\strucs}$ from $\strucs$.

\skipnoindent \underline{Base case:} Assume $\astruc \in \strucs$.
Then $\astruc' = \astruc$ satisfies the property.

\skipnoindent \underline{Induction step:} Assume $\astruc = (\astruc_1
\comp \astruc_2)_{/\approx}$ for some $\astruc_1, \astruc_2 \in
\fclose{}{\strucs}$ and some equivalence relation $\approx$ generated
by the independent set $\set{(u_{1i}, u_{2i}) \mid i \in
  \interv{1}{n}}$, that conforms to the compatibility condition of
Definition \ref{def:fusion} for $\astruc_1, \astruc_2$. Let
$\acolor_{1i} = \colorof{}{\astruc_1}(u_{1i})$, $\acolor_{2i} =
\colorof{}{\astruc_2}(u_{2i})$, for all $i\in\interv{1}{n}$. By
Definition \ref{def:fusion}, $\astruc = (\astruc_1 \comp
\astruc_2)_{/\approx}$ implies $\acolor_{1i} \cap \acolor_{2i} =
\emptyset$, for all $i \in \interv{1}{n}$, and moreover:
\[
  \mcolabs{\astruc} = \mset{(\acolor_{1i} \cup \acolor_{2i}) \mid i\in\interv{1}{n}} \cup
  (\mcolabs{\astruc_1} \setminus \mset{(\acolor_{1i}) \mid i\in\interv{1}{n}}) \cup
  (\mcolabs{\astruc_2} \setminus \mset{(\acolor_{2i}) \mid i\in\interv{1}{n}})
\]
By the inductive hypothesis, there exists
$\astruc_1', \astruc_2' \in \fclose{1}{\strucs}$ such that
$\mcolabs{\astruc_1} \subseteq \mcolabs{\astruc_1'}$,
$\mcolabs{\astruc_2} \subseteq \mcolabs{\astruc_2'}$. We use
$\astruc_1'$ and $n$ disjoint copies $\astruc_{2,1}', ...,
\astruc_{2,n}'$ of $\astruc_2'$ to construct $\astruc'$ with the
required property. The idea is that, for every pair $u_{1i} \approx
u_{2i}$, we fuse some element $u'_{1i}$ with color $\acolor_{1i}$ from
$\astruc_1'$ with some element $u'_{2i}$ with color $\acolor_{2i}$
from $\astruc_{2,i}'$. Such elements always exist, because
$\mcolabs{\astruc_1} \subseteq \mcolabs{\astruc_1'}$,
$\mcolabs{\astruc_2} \subseteq \mcolabs{\astruc_2'}$. Therefore,
consider the equivalence relations $\approx_i'$ generated by $\set{u_{1i}', u_{2i}'}$
for some pair of elements as above, for each $i\in\interv{1}{n}$, and
define:
\[
  \astruc' \isdef ( \ldots ((\astruc_1' \comp \astruc_{2,1}')_{/\approx'_1} \comp
  \astruc_{2,2}')_{/\approx'_2} \comp \ldots \comp \astruc_{2,n}')_{\approx'_n}
\]
Then $\astruc' \in \fclose{1}{\strucs}$ and, moreover, we have
$\mcolabs{\astruc} \subseteq \mcolabs{\astruc'}$, because:
\[
  \mcolabs{\astruc'} = \mset{(\acolor_{1i} \cup \acolor_{2i}) \mid i\in\interv{1}{n}} \cup
  (\mcolabs{\astruc_1'} \setminus \mset{(\acolor_{1i}) \mid i\in\interv{1}{n}}) \cup
  \bigcup\nolimits_{i\in\interv{1}{n}} (\mcolabs{\astruc_2'} \setminus \mset{\acolor_{2i}}) \hspace{1cm} \qedhere
\]
\end{proofE}

The set $\kmcolabs{k}{(\fclose{1}{\strucs})}$ can be computed by a
least fixpoint iteration of the following abstract operation on the
domain of $k$-multiset color abstractions. As the later domain is
finite, this fixpoint computation is guaranteed to terminate.

\begin{definition}\label{def:multiset-color-fusion-abstraction}
  The \emph{single-pair multiset fusion} is defined below, for all $M_1, M_2 \in \mpow{\allcols}$:
  \[\begin{array}{rl}
    \absextfusionone{M_1}{M_2} \isdef
    \big\{M \in \mpow{\allcols} \mid & \exists \acolor_1 \in M_1 ~.~ \exists \acolor_2\in M_2 ~.~ \acolor_1 \cap \acolor_2 = \emptyset, \\
    & ~M = \mset{\acolor_1 \cup \acolor_2} \cup \bigcup\nolimits_{i=1,2} (M_i \setminus \mset{\acolor_i}) \big\}
  \end{array}\]
  Given an integer $k\geq1$, the \emph{single-pair $k$-multiset
    fusion} is defined for $M_1$, $M_2 \in \mpow{\allcols}$, such that
  $\cardof{M_1} \le k$ and $\cardof{M_2}\le k$:
  \[
  \kabsextfusionone{k}{M_1}{M_2} \isdef \set{M ~|~ \exists M' \in
    \absextfusionone{M_1}{M_2}.~ M \subseteq M',~ \cardof{M} \le k}
  \]
  For a set $\mathcal{M}$ of multisets (resp. $k$-multisets) of
  colors, let $\absreachfusionone{{\mathcal M}}$
  (resp. $\kabsreachfusionone{k}{\mathcal M}$) be the closure of
  $\mathcal{M}$ under taking single-pair fusion on multisets
  (resp. $k$-multisets).
\end{definition}
This operation is used to compute
$\kmcolabs{k}{(\fclose{1}{\strucs})}$ by a iterating
$\kabsreachfusionone{k}{}$ starting with $\kmcolabs{k}{\strucs}$ until
a fixed point is reached. Since there are finitely many colors, the
domain of multisets of colors having multiplicity at most $k$ is
finite, hence this iteration is guaranteed to compute
$\kabsreachfusionone{k}{\kmcolabs{k}{\strucs}}$ in finitely many
steps.

\begin{lemmaE}\label{lemma:reach-fusion-one-abstraction}
  $\kmcolabs{k}{(\fclose{1}{\strucs})} =
  \kabsreachfusionone{k}{\kmcolabs{k}{\strucs}}$, for any set
  $\strucs$ of structures and integer $k\geq1$.
\end{lemmaE}
\begin{proofE}
    Abusing notation, we write $\kmcolabs{k}{M} \isdef \set{M' ~|~ M'
    \subseteq M,~ \cardof{M'} \le k}$. Then, we have
  $\kmcolabs{k}{(\fclose{1}{\strucs})} =
  \kmcolabs{k}{(\mcolabs{(\fclose{1}{\strucs})})}$, by
  \autoref{def:multiset-color-abstraction}. We can prove that for all structures
  $\astruc_1, \astruc_2$ it holds
  $\mcolabs{(\extfusionone{\astruc_1}{\astruc_2})} =
  \absextfusionone{\mcolabs{\astruc_1}}{\mcolabs{\astruc_2}}$. This
  immediately extends to their respective closure, henceforth,
  $\mcolabs{(\fclose{1}{\strucs})} =
  \absreachfusionone{\mcolabs{\strucs}}$. Henceforth, we are left
  with proving that
  $\kmcolabs{k}{(\absreachfusionone{\mcolabs{\strucs}})} =
  \kabsreachfusionone{k}{\kmcolabs{k}{\strucs}}$.

  \skipnoindent ``$\kmcolabs{k}{(\absreachfusionone{\mcolabs{\strucs}})}
  \subseteq \kabsreachfusionone{k}{\kmcolabs{k}{\strucs}}$'' We
  prove that, for all $M \in
  \absreachfusionone{\mcolabs{\strucs}}$, we have $\kmcolabs{k}{M}
  \subseteq \kabsreachfusionone{k}{\kmcolabs{k}{\strucs}}$. The
  proof goes by induction on the derivation of $M\in
  \absreachfusionone{\mcolabs{\strucs}}$ from
  $\mcolabs{\strucs}$.

  \skipnoindent \underline{Base case:} Assume $M
  \in \mcolabs{\strucs}$.  Then $\kmcolabs{k}{M} \subseteq
  \kmcolabs{k}{(\mcolabs{\strucs})} = \kmcolabs{k}{\strucs}
  \subseteq \kabsreachfusionone{k}{\kmcolabs{k}{\strucs}}$.

  \skipnoindent \underline{Induction step:} Assume $M \in
  \absextfusionone{M_1}{M_2}$ for some multisets of colors $M_1$,
  $M_2$ such that $\kmcolabs{k}{M_1}, \kmcolabs{k}{M_2} \subseteq
  \kabsreachfusionone{k}{\kmcolabs{k}{\strucs}}$. Then, there
  exists $\acolor_1 \in M_1$, $\acolor_2 \in M_2$ such that $\acolor_1
  \cap \acolor_2 = \emptyset$ and $M = (M_1 \setminus
  \mset{\acolor_1}) \cup (M_2 \setminus \mset{\acolor_2}) \cup
  \mset{\acolor_1 \cup \acolor_2}$, by
  \autoref{def:multiset-color-fusion-abstraction}. Let $M' \in
  \kmcolabs{k}{M}$, that is, $M' \subseteq M$, $\cardof{M'} \le k$.
  We distinguish several cases: \begin{itemize}
  \item $M' \subseteq M_1$ (the case $M' \subseteq M_2$ is symmetric):
    $M' \in \kmcolabs{k}{M_1}$, thus $M' \in
    \kabsreachfusionone{k}{\kmcolabs{k}{\strucs}}$.
  \item $M' \not\subseteq M_i$, for $i=1,2$ and $\acolor_1 \cup
    \acolor_2 \not\in M$': $M'$ can be partitioned in two nonempty
    parts $M_1' \subseteq M_1$, $M_2' \subseteq M_2$ such that $M =
    M_1' \uplus M_2'$. As both parts are not empty, we have $M_1' \in
    \kmcolabs{k-1}{M_1}$, $M_2' \in \kmcolabs{k-1}{M_2}$, thus $(M_1'
    \cup \mset{\acolor_1}) \in \kmcolabs{k}{M_1}$, $(M_2' \cup
    \mset{\acolor_2}) \in \kmcolabs{k}{M_2}$. It is an easy check that
    $M' \in \kabsextfusionone{k}{(M_1' \cup \mset{\acolor_1})}{(M_2'
      \cup \mset{\acolor_2})}$. This implies $M' \in
    \kabsreachfusionone{k}{\kmcolabs{k}{\strucs}}$ as both
    subterms belong to
    $\kabsreachfusionone{k}{\kmcolabs{k}{\strucs}}$.
  \item $M' \not\subseteq M_i$, for $i=1,2$ and $\acolor_1 \cup
    \acolor_2 \in M'$: we proceed as in the previous case but
    considering a partitioning of $M' \setminus \mset{\acolor_1 \cup
      \acolor_2}$.  We obtain $M' \in
    \kabsreachfusionone{k}{\kmcolabs{k}{\strucs}}$, as well.
  \end{itemize}

  \skipnoindent ``$\kabsreachfusionone{k}{\kmcolabs{k}{\strucs}} \subseteq
    \kmcolabs{k}{(\absreachfusionone{\mcolabs{\strucs}})}$'' We
  prove that, for all $k$-multiset $M'\in
  \kabsreachfusionone{k}{\kmcolabs{k}{\strucs}}$, there exists $M \in
  \absreachfusionone{\mcolabs{\strucs}}$, such that $M' \subseteq M$, by
  induction on the derivation of $M'$ from $\kmcolabs{k}{\strucs}$.

  \skipnoindent \underline{Base case:} Assume $M'
  \in \kmcolabs{k}{\strucs} =
  \kmcolabs{k}{(\mcolabs{\strucs})}$. Then, there exists $M\in
  \mcolabs{S}$ such that $M' \subseteq M$.  Obviously, $M \in
  \absreachfusionone{\mcolabs{\strucs}}$.

  \skipnoindent \underline{Induction step:} Assume $M' \in
  \kabsextfusionone{k}{M_1'}{M_2'}$ for some $k$-multisets of colors
  $M_1', M_2' \in \kabsreachfusionone{k}{\kmcolabs{k}{\strucs}}$.
  By the inductive hypothesis, there exists multisets $M_1, M_2 \in
  \absreachfusionone{\mcolabs{\strucs}}$ such that $M_1' \subseteq
  M_1$, $M_2' \subseteq M_2$.  Since $M_1', M_2'$ can be composed such
  that to obtain (a superset of) the multiset $M'$, one can use
  precisely the same pairs of colors to compose $M_1, M_2$ and
  henceforth to obtain the multiset $M \in
  \absreachfusionone{\mcolabs{\strucs}}$, which is the superset of
  $M'$.
\end{proofE}

The last step concerns the effective computation of
$\kmcolabs{k}{\strucs}$, i.e., the set of color multisets of
multiplicity at most $k$ that occur in the multiset abstraction of
some structure $\astruc \in \strucs$. We leverage from the fact that
$\strucs$ is a context-free set described by the set of terms that
forms the language of a tree automaton $\mathcal{A}$.

We assume basic acquaintance with \emph{context-free grammars}, i.e.,
finite sets of rules of the form either $q \leftarrow
t[q_1,\ldots,q_k]$ or $\leftarrow q$, where $q,q_1,\ldots,q_k$ denote
nonterminals and $t$ is a $\fsignature_\algof{HR}$-term with variables
$q_1,\ldots,q_k$. The language $\lang{\grammar}$ of a grammar
$\grammar$ is the set of interpretations in $\algof{HR}$ of the terms
produced by derivations starting with an axiom $\leftarrow q$. It is
well known that a tree automaton can be transformed into a
context-free grammar having the same language, by turning each
transition $(q_1,\ldots,q_k) \arrow{f}{} q$ into a rule $q \leftarrow
f[q_1,\ldots,q_k]$, for $k\geq1$, and adding an axiom $\leftarrow q$
for each final state $q$ of the tree automaton.

By \emph{first-order logic} we understand the set of formul{\ae}
consisting of equalities between variables, relation atoms of the form
$\arel(x_1,\ldots,x_{\arityof{\arel}})$, for some relation symbol
$\arel$, composed via boolean operations and quantifiers. A
first-order logic sentence $\varphi$ (i.e., a formula without free
variables) is interpreted over a structure $\astruc$ by the
satisfiability relation $\astruc \models \varphi$, defined inductively
on the structure of $\varphi$, as usual.


Over words, it is a well-known result that the non-emptiness of the
intersection of a context-free set (e.g., given by a context-free
grammar) and a regular set (e.g., given by a regular grammar, DFA, or
a MSO formula) is decidable.  This result has been generalized by
Courcelle to context-free grammars over the \hr-algebra of structures
and FO-definable sets of structures\footnote{\cite[Theorem
    3.6]{CourcelleVII} is actually given for Monadic Second Order
  Logic, which subsumes first-order logic.}:

\begin{theorem}[Theorem 3.6 in \cite{CourcelleVII}]\label{thm:ft}
  For each grammar $\grammar$ and first-order sentence $\varphi$, one
  can decide the existence of a structure $\astruc \in
  \lang{\grammar}$ such that $\astruc \models \varphi$.
\end{theorem}

In order to compute $\kmcolabs{k}{\strucs}$, we build a first-order
sentence $\varphi_M$ for each multiset of colors $M : \allcols
\rightarrow \interv{1}{k}$ such that $\astruc \models \varphi_M$ iff
$\kmcolabs{k}{\astruc}$. As there are only finitely many such
multi-sets $M$, we are able to construct $\kmcolabs{k}{\strucs}$ by
finitely many calls to the above decision procedure.  We now state the
details for the construction of $\varphi_M$.  For each color $\acolor
\in \allcols$, we denote by $\varphi_\acolor(X)$ the formula
$\bigwedge_{\arel \in \colors} \arel(x) \wedge \bigwedge_{\arel
  \not\in \colors} \neg \arel(x)$.  We then obtain $\varphi_M$ as the
conjunction of the formul{\ae} $\exists^{=M(\acolor)} x ~.~
\varphi_\acolor(x)$, if $M(\acolor) < k$, and $\exists^{\ge
  M(\acolor)} x ~.~ \varphi_\acolor(x)$, if $M(\acolor) = k$, for all
colors $\acolor \in \allcols$. As usual, the quantifier $\exists^{=n}
x ~.~ \phi(x)$ (resp. $\exists^{\geq n} x ~.~ \phi(x)$) means ``there
exists exactly $n$ (resp. at least $n$) elements $x$ that satisfy
$\phi(x)$''. It is now easy to verify that $\astruc \models \varphi_M$
iff $M \in \kmcolabs{k}{\astruc}$.  This concludes the proof of
Theorem~\ref{thm:main} (\ref{it2:thm:main}).

\section{Conclusions and Future Work}

We have defined a non-aggregative and nondeterministic fusion
operation on logical structures, that is controlled by a coloring of
structures using unary relations. We study the tree-width of the
closure of a context-free set under fusion. We prove that it is
decidable whether the closure of a context-free set has bounded
tree-width. Moreover, if this is the case, we show that the closure
set is context-free as well, described by an effectively constructible
grammar.

Future work involves considering more general notions of coloring,
e.g., coloring functions defined by \mso-definable
transductions. Moreover, we plan to investigate generalizations of
Theorem \ref{thm:main} for other notions of width, such as
generalizations of clique-width and rank-width from graphs and
hypergraphs to relational structures.

\bibliography{refs}

\begin{thebibliography}{10}

\bibitem{AhrensBozgaIosifKatoen21}
Emma Ahrens, Marius Bozga, Radu Iosif, and Joost{-}Pieter Katoen.
\newblock Reasoning about distributed reconfigurable systems.
\newblock {\em Proc. {ACM} Program. Lang.}, 6({OOPSLA2}):145--174, 2022.
\newblock \href {https://doi.org/10.1145/3563293} {\path{doi:10.1145/3563293}}.

\bibitem{conf/popl/CalcagnoGZ05}
Cristiano Calcagno, Philippa Gardner, and Uri Zarfaty.
\newblock Context logic and tree update.
\newblock In Jens Palsberg and Mart{\'{\i}}n Abadi, editors, {\em Proceedings
  of the 32nd {ACM} {SIGPLAN-SIGACT} Symposium on Principles of Programming
  Languages, {POPL} 2005, Long Beach, California, USA, January 12-14, 2005},
  pages 271--282. {ACM}, 2005.
\newblock URL: \url{https://doi.org/10.1145/1040305.1040328}.

\bibitem{conf/popl/CalcagnoGZ07}
Cristiano Calcagno, Philippa Gardner, and Uri Zarfaty.
\newblock Context logic as modal logic: completeness and parametric
  inexpressivity.
\newblock In Martin Hofmann and Matthias Felleisen, editors, {\em Proceedings
  of the 34th {ACM} {SIGPLAN-SIGACT} Symposium on Principles of Programming
  Languages, {POPL} 2007, Nice, France, January 17-19, 2007}, pages 123--134.
  {ACM}, 2007.
\newblock \href {https://doi.org/10.1145/1190216.1190236}
  {\path{doi:10.1145/1190216.1190236}}.

\bibitem{Cardelli2002Spatial}
Luca Cardelli, Philippa Gardner, and Giorgio Ghelli.
\newblock {A Spatial Logic for Querying Graphs}.
\newblock In Peter Widmayer, Francisco~Triguero Ruiz, Rafael~Morales Bueno,
  Matthew Hennessy, Stephan Eidenbenz, and Ricardo Conejo, editors, {\em
  Proceedings of the 29\textsuperscript{th} International Colloquium on
  Automata, Languages and Programming ({ICALP'02})}, volume 2380 of {\em
  Lecture Notes in Computer Science}, pages 597--610. Springer, July 2002.

\bibitem{comon:hal-03367725}
Hubert Comon, Max Dauchet, R{\'e}mi Gilleron, Florent Jacquemard, Denis Lugiez,
  Christof L{\"o}ding, Sophie Tison, and Marc Tommasi.
\newblock {\em {Tree Automata Techniques and Applications}}.
\newblock HAL, 2008.
\newblock URL: \url{https://inria.hal.science/hal-03367725}.

\bibitem{CourcelleI}
Bruno Courcelle.
\newblock The monadic second-order logic of graphs. {I}. {R}ecognizable sets of
  finite graphs.
\newblock {\em Information and Computation}, 85(1):12--75, 1990.
\newblock \href {https://doi.org/10.1016/0890-5401(90)90043-H}
  {\path{doi:10.1016/0890-5401(90)90043-H}}.

\bibitem{CourcelleVII}
Bruno Courcelle.
\newblock The monadic second-order logic of graphs {VII:} graphs as relational
  structures.
\newblock {\em Theor. Comput. Sci.}, 101(1):3--33, 1992.
\newblock \href {https://doi.org/10.1016/0304-3975(92)90148-9}
  {\path{doi:10.1016/0304-3975(92)90148-9}}.

\bibitem{courcelle_engelfriet_2012}
Bruno Courcelle and Joost Engelfriet.
\newblock {\em Graph Structure and Monadic Second-Order Logic: A
  Language-Theoretic Approach}.
\newblock Encyclopedia of Mathematics and its Applications. Cambridge
  University Press, 2012.
\newblock \href {https://doi.org/10.1017/CBO9780511977619}
  {\path{doi:10.1017/CBO9780511977619}}.

\bibitem{journals/lmcs/DochertyP18}
Simon Docherty and David~J. Pym.
\newblock Intuitionistic layered graph logic: Semantics and proof theory.
\newblock {\em Log. Methods Comput. Sci.}, 14(4), 2018.
\newblock \href {https://doi.org/10.23638/LMCS-14(4:11)2018}
  {\path{doi:10.23638/LMCS-14(4:11)2018}}.

\bibitem{DBLP:books/daglib/0082516}
Heinz{-}Dieter Ebbinghaus and J{\"{o}}rg Flum.
\newblock {\em Finite model theory}.
\newblock Perspectives in Mathematical Logic. Springer, 1995.

\bibitem{DBLP:series/txtcs/FlumG06}
J{\"{o}}rg Flum and Martin Grohe.
\newblock {\em Parameterized Complexity Theory}.
\newblock Texts in Theoretical Computer Science. An {EATCS} Series. Springer,
  2006.
\newblock \href {https://doi.org/10.1007/3-540-29953-X}
  {\path{doi:10.1007/3-540-29953-X}}.

\bibitem{Concur23}
Radu Iosif and Florian Zuleger.
\newblock Expressiveness results for an inductive logic of separated relations.
\newblock In Guillermo~A. P\'{e}rez and Jean-Fran\c{c}ois Raskin, editors, {\em
  34th International Conference on Concurrency Theory (CONCUR 2023)}, volume
  279 of {\em Leibniz International Proceedings in Informatics (LIPIcs)}, pages
  20:1--20:20, Dagstuhl, Germany, 2023. Schloss Dagstuhl -- Leibniz-Zentrum
  f{\"u}r Informatik.
\newblock \href {https://doi.org/10.4230/LIPIcs.CONCUR.2023.20}
  {\path{doi:10.4230/LIPIcs.CONCUR.2023.20}}.

\bibitem{Ishtiaq00bias}
Samin~S. Ishtiaq and Peter~W. O'Hearn.
\newblock {BI} as an assertion language for mutable data structures.
\newblock In Chris Hankin and Dave Schmidt, editors, {\em Conference Record of
  {POPL} 2001: The 28th {ACM} {SIGPLAN-SIGACT} Symposium on Principles of
  Programming Languages, London, UK, January 17-19, 2001}, pages 14--26. {ACM},
  2001.

\bibitem{10.1007/978-3-540-27864-1_26}
Viktor Kuncak and Martin Rinard.
\newblock Generalized records and spatial conjunction in role logic.
\newblock In {\em Static Analysis}, pages 361--376, Berlin, Heidelberg, 2004.
  Springer Berlin Heidelberg.

\bibitem{OHearnReynoldsYang01}
Peter~W. O'Hearn, John~C. Reynolds, and Hongseok Yang.
\newblock Local reasoning about programs that alter data structures.
\newblock In {\em Proceedings of the 15th International Workshop on Computer
  Science Logic}, CSL '01, pages 1--19, 2001.

\bibitem{PYM2004257}
David~J. Pym, Peter~W. O'Hearn, and Hongseok Yang.
\newblock Possible worlds and resources: the semantics of bi.
\newblock {\em Theoretical Computer Science}, 315(1):257--305, 2004.
\newblock Mathematical Foundations of Programming Semantics.
\newblock URL:
  \url{https://www.sciencedirect.com/science/article/pii/S0304397503006248},
  \href {https://doi.org/https://doi.org/10.1016/j.tcs.2003.11.020}
  {\path{doi:https://doi.org/10.1016/j.tcs.2003.11.020}}.

\bibitem{sep-logic-substructural}
Greg Restall.
\newblock {Substructural Logics}.
\newblock In Edward~N. Zalta and Uri Nodelman, editors, {\em The {Stanford}
  Encyclopedia of Philosophy}. Metaphysics Research Lab, Stanford University,
  {F}all 2024 edition, 2024.

\bibitem{Reynolds02}
John~C. Reynolds.
\newblock Separation logic: {A} logic for shared mutable data structures.
\newblock In {\em 17th {IEEE} Symposium on Logic in Computer Science {(LICS}
  2002), 22-25 July 2002, Copenhagen, Denmark, Proceedings}, pages 55--74.
  {IEEE} Computer Society, 2002.
\newblock \href {https://doi.org/10.1109/LICS.2002.1029817}
  {\path{doi:10.1109/LICS.2002.1029817}}.

\bibitem{Seese91}
D.~Seese.
\newblock The structure of the models of decidable monadic theories of graphs.
\newblock {\em Annals of Pure and Applied Logic}, 53(2):169--195, 1991.
\newblock \href {https://doi.org/10.1016/0168-0072(91)90054-P}
  {\path{doi:10.1016/0168-0072(91)90054-P}}.

\bibitem{SEYMOUR199322}
P.D. Seymour and R.~Thomas.
\newblock Graph searching and a min-max theorem for tree-width.
\newblock {\em Journal of Combinatorial Theory, Series B}, 58(1):22--33, 1993.
\newblock \href {https://doi.org/10.1006/jctb.1993.1027}
  {\path{doi:10.1006/jctb.1993.1027}}.

\end{thebibliography}

\appendix
\section{Proofs}
\label{app:proofs}
\printProofs

\end{document}